\documentclass[preprint,3p,times, 12pt]{elsarticle}

\usepackage{amssymb}
\usepackage{amsmath}
\usepackage{booktabs}
\usepackage[hidelinks]{hyperref}
\usepackage{xcolor}
\usepackage{xcolor}
\usepackage{threeparttable}
\allowdisplaybreaks

\journal{Acta Materialia}

\begin{document}

\begin{frontmatter}



\title{A mechanistic model for vacancy energetics in concentrated solid solutions with short-range order}


\author[inst1]{Akash Baski} 
\author[inst1]{Ritesh Jagatramka}
\author[inst1]{Matthew Daly\corref{cor1}}
\cortext[cor1]{Corresponding author}
\ead{mattdaly@uic.edu}
\affiliation[inst1]{organization={Department of Civil, Materials, and Environmental                     Engineering, University of Illinois Chicago},
            addressline={842 W. Taylor St., 2095 ERF (MC 246)}, 
            city={Chicago},
            postcode={60607}, 
            state={Illinois},
            country={United States}}

\begin{abstract}
Understanding the vacancy-mediated properties of concentrated solid solutions (CSSs) is essential to predict their mechanical behavior under diffusional flow. However, the metallurgy of the underlying vacancy energetics remains challenging to treat due to the variable chemical environment inherent to this class of alloys. Here, we present a mechanistic model to predict the vacancy energy distributions of CSSs with an arbitrary number of components, composition, and short-range order (SRO). This model possesses no adjustable parameters and consists of a series of closed-form analytical relations that can be calculated with trivial computational effort. Although several approximations are necessary to achieve generality, predictions from the model rival the resource-intensive methods used in common practice in the systems studied. Guided by the mechanistic model, we explore the sensitivities of the vacancy-mediated properties of CSSs with different levels of SRO. We offer this model for rapid computational screening of the CSS design space and to accelerate alloy design efforts. Computational scripts to execute the key relations of the mechanistic model have been made available to the community through an online code repository.
\end{abstract}

%


\end{frontmatter}




\section{Introduction}

Compared with traditional dilute alloys, concentrated solid solutions (CSS) exhibit distinctive defect behaviors, which emerge from their variable chemical arrangement. Here, we classify CSSs as solid solutions where multiple elements (two or more) are found in non-dilute proportions as a single-phase mixture. This definition shares characteristics with other nomenclature familiar to the metallurgy community, including classifiers that specify entropy stabilization (i.e., medium and high entropy alloys), and broader umbrella terms that relax the single-phase constraint (i.e., multi-principal element and complex concentrated alloys). Several high-quality review articles are available that describe the metallurgy of CSSs and the broader families of complex alloys~\cite{zhang2014microstructures, miracle2017critical, george2019high, george2020high}. In CSSs, the impact of chemical arrangement on defect properties becomes significant when the length-scale of the defect structure and chemical heterogeneity overlap, causing significant metallurgical interaction~\cite{jagatramka2023evolution}. Within these length-scale considerations, vacancies serve as an important lattice defect governing diffusional flow, with the effects of variable chemical arrangement on vacancy formation energies (VFE) and migration energies (VME) well documented~\cite{Zhao2017,Osetsky2016,23,24}. In contrast to studies of pure metals and dilute alloys, a general observation from these investigations is that VFEs and VMEs are distributed rather than single-valued. Consequently, these vacancy energy fluctuations significantly impact defect kinetics following primary knock-on atom (PKA) damage~\cite{deluigi2021simulations, 34, Beland2016, zhao2017suppression}, as well as atomic-scale mass transport~\cite{Osetsky2018,xu2023mechanism,xing2024vacancy}. 

Atomistic and first-principles investigations have attempted to elucidate how local atomic coordination governs these distributions. For instance, Zhao et al. demonstrated that chemical disorder in Ni$_{0.5}$Fe$_{0.5}$ and Ni$_{0.8}$Fe$_{0.2}$ alloys gives rise to broad VFE profiles, wherein locally Ni-enriched environments lower the Ni VFE by up to $\sim0.5$~eV~\cite{23}. Similar configuration-dependent VFEs have been systematically characterized across a wide range of face-centered cubic (FCC) and some hexagonal close-packed alloys~\cite{li2019first, kang2023density, 35, zhang2022ab}. In parallel, the local chemical environment is also known to strongly modulate the vacancy migration energies. First-principles calculations by Gao et al. in the Cantor alloy (CoCrFeMnNi) revealed a broad VME distribution comprised of species-specific sub-spectra~\cite{gao2024defect} and Zhao reported VMEs spanning the remarkably wide range of 0.07--1.95~eV in refractory VTaCrW alloys~\cite{zhao2020defect}. Moreover, a positron annihilation spectroscopy study from An et al.~\cite{32} showed Al addition in proton-irradiated $\text{Al}_x\text{FeCoCrNi}$ substantially accelerates vacancy recovery, providing experimental observation of the solute effect. Combined with nudged elastic band calculations, these investigators linked the increased recovery rate with shifts in the VME distribution, where vacancies exhibited preferential annihilation within Al-rich environments. 

The distributions in vacancy energies measured in CSSs are intimately connected to the patterns of atom arrangement that form the chemical structure. Although many early studies have focused on random arrangement, chemical correlations such as those produced by short-range order (SRO) have become a topic of recent interest~\cite{kang2023density, rasooli2024deformation, cao2021does, xing2022short}. Here, SRO emerges from thermodynamically preferred chemical arrangements within one or more lattice neighbor shells, thereby altering the statistical distribution of atomic coordination. These structural correlations directly affect the local chemical environment surrounding a vacancy, leading to shifts in the VFE and VME~\cite{li2024vacancy,manzoor2024interplay}. Relative to random mixing, the preferential chemical correlations introduced by SRO are also reported to shift tendencies for forward and backward migration. That is, by modifying the local chemical environment, SRO changes not only the distributions of the migration barriers but also introduces spatial correlations into the energy landscape. Consequently, SRO gives rise to correlated diffusion pathways that result in marked deviations from random-walk models~\cite{xing2022short, xu2023influence, xing2024neural}.
 
While the effects of varied chemical arrangement on vacancies and their associated metallurgical processes have been documented, mechanistic models that connect the chemical structure of CSSs with the emergent defect energies remain underdeveloped. Current practice leverages computational tools such as density functional theory (DFT) and molecular dynamics simulations to directly calculate vacancy energies. For instance, Zhang et al.~\cite{27} developed a workflow to calculate the VFE statistics of randomly arranged CSSs from atomistic calculations, but their method cannot be readily extended to systems with SRO because it relies on assumptions that hold only for random arrangement. More recently, Lee et al.~\cite{Lee2026} studied the equilibrium vacancy concentration of CSSs by combining DFT with cluster expansion and statistical mechanics methods to capture chemical ordering effects, but at substantial computational cost. Moreover, although these direct computational approaches provide high accuracy and fidelity, the vast configurational space offered by CSSs presents large bottlenecks for computational screening, with tools to rapidly examine the underlying structure-property relationships remaining largely absent. Machine-learning-based approaches have emerged as promising alternatives to accelerate the prediction of vacancy formation and migration energetics~\cite{ibrahim2026machine,fan2022predicting,castin2014predicting}. Multiple recent studies have demonstrated moderate-to-high predictive accuracy when models are trained on sufficiently large atomistic datasets and employ carefully engineered structural descriptors or graph-based representations. However, these approaches remain intrinsically dependent on extensive training databases generated via DFT or other atomistic simulations, the construction of which becomes prohibitively expensive for complex alloys with large configurational spaces. Furthermore, as machine learning models typically capture statistical correlations rather than explicitly encoding the underlying defect physics, their transferability across alloy compositions, chemical ordering states, and previously unexplored material systems remains a significant and unresolved challenge. Although the literature offers some examples of machine learning methods to predict vacancy energies from the available defect-free DFT databases, their error is large. For example, a mean absolute error of 1.19 eV has been reported for pure element VFEs predicted under such methods~\cite{choudhary2023can}, which approaches the nominal values for most metals (e.g., $\sim$1.5~eV in pure Ni). 

To address these shortcomings, we present a mechanistic model to predict the distributions of vacancy energies in CSSs under both random arrangement and with SRO. Our approach is grounded in a statistical analysis of chemical arrangement during vacancy formation and migration, with defect energy predictions emerging from an analysis of excess energy distributions in the vicinity of the vacancy. Our model is analytical and only requires the input of known defect structure factors and interaction energies that are available directly from interatomic potentials. The vacancy energy predictions are therefore high throughput and extendable to systems of arbitrary composition, components, and arrangements. We assess the accuracy of our model predictions through direct simulation across several benchmarking CSSs. Furthermore, we leverage our predictions for vacancy energy distributions to assess the effect of defect statistics on the relevant defect-mediated properties (i.e., vacancy concentration and vacancy diffusivity). For this purpose, we develop an effective barrier model to account for thermal activation of reduced barriers available at the distribution tail. We compare our effective barrier property predictions to single-valued traditional models and validate our approach using direct simulation and literature data, where available. We have also written scripts to facilitate adoption of this method within the community. These tools can be accessed through a GitHub repository~\cite{AdvancedMaterialsandMicrostructuresLabattheUniversityofIllinoisChicago}.

\section{Mechanistic model}\label{Anal}

\subsection{Cohesive energy statistics in CSSs with short-range order}\label{Theory}

Our approach to predict vacancy energy distributions in CSSs begins from consideration of cohesive energy, which we later demonstrate to be a fundamental input of our model. As with vacancy energies, the cohesive energy also exhibits a statistical distribution in CSSs. This distribution can be taken as unimodal and Gaussian~\cite{jagatramka2022analytical} and is therefore fully defined by its mean and standard deviation. To evaluate these quantities, we ground our approach in the method described by Varvenne et al.~\cite{varvenne2016average}, where energies are homogenized to produce an average-atom interatomic potential for randomly arranged CSSs under the embedded atom method (EAM)~\cite{38, 39, 40}. In the standard EAM formulation, the potential energy is expressed as a system-level state variable that requires summation over all atomic sites and chemical species. Our previous work~\cite{jagatramka2022analytical} recast system-level, average-atom expressions to provide energies in specific terms (i.e., as cohesive energies). This approach leverages structure factors fixed by lattice symmetry to treat the chemical environment, from which the statistics of cohesive energies are defined by the probability of atom placement at discrete distances. The mean cohesive energy ($\bar{e}$) for a randomly arranged alloy is the sum of the mean embedding energy ($\bar{F}$) and mean pair interaction energy ($\bar{\phi}$)~\cite{jagatramka2022analytical}:
\begin{equation}
\label{eq:ecoh}
    \bar{e} = \overbrace{\sum_{X} c^{X} F^{X}(\bar{\rho})}^{\bar{F}} + \frac{1}{2}\overbrace{\sum_{X,Y,\zeta} {c}^{X}{c}^{Y} Z_{\zeta} \phi_{\zeta}^{XY}}^{\bar{\phi}}  
\end{equation}
with $XY$ defining interacting chemical pairs, $c$ as the molar fractional composition, $Z$ as the structure factor, and $\bar{\rho}$ as the mean per-atom electron density. The structure factor is described here as the atomic coordination within successive neighbor shells ($\zeta$) enumerated radially outward from a central atom. For example, in the case of an ideal FCC lattice, the first neighbor shell ($\zeta$ = 1) has a structure factor of $Z_1$ = 12. The term $\phi_\zeta^{XY}$ represents the interaction energies between atoms $XY$ at distances matching the neighbor shell $\zeta$. The mean per-atom electron density is given as:
\begin{equation}
\label{eq:rho_rand}
    \bar{\rho} = \sum_{Y,\zeta} c^Y Z_{\zeta} \rho_{\zeta}^Y 
\end{equation}
with $\rho_{\zeta}^Y$ as the electron density contributed by atom $Y$ within the coordination shell $\zeta$. Although this derivation proceeds from the EAM formulation, we note that the relations are general to other pair-functional potentials, and all pair potentials, provided interaction energies can be parameterized by structure factors. While mathematically simple, DFT-fit pair potentials have proven accurate in modeling local defect energy fluctuations in CSSs with SRO~\cite{LaRosa2022}.

To capture the effects of SRO on cohesive energies, we extend our previous expressions for random arrangement to account for preferential chemical correlations and then propagate the new relations through the statistical considerations undertaken in Jagatramka et al.~\cite{jagatramka2022analytical}. As a two-body descriptor, the Warren-Cowley (WC) parameter~\cite{cowley1950x,cowley1965short} is used to parameterize chemical correlations in the EAM formulation. Although higher-order chemical motifs are required to uniquely describe the topology of CSSs~\cite{Sheriff2024}, our results show that the effects of SRO on cohesive energy distributions are sufficiently captured by pair correlations. The WC parameter may be equivalently written using a conditional or joint probability definition as:
\begin{equation}
\label{eq:WC}
    \alpha_{\zeta}^{XY} = 1- \frac{p_{\zeta}^{Y|X}}{c^{Y}} = 1 - \frac{p_{\zeta}^{XY}}{c^{X}c^{Y}} 
\end{equation}
where $p_{\zeta}^{Y|X}$ is the conditional probability of finding atom $Y$ in neighbor shell $\zeta$ around atom $X$ and $p_{\zeta}^{XY}$ is the joint probability of the same pairing, with normalization of the conditional probabilities requiring $\sum_{Y}c^{Y}\alpha_{\zeta}^{XY} = 0$. Recognizing that the $c^{X}c^{Y}$ and $c^{Y}$ terms in Eqs.~\ref{eq:ecoh} and~\ref{eq:rho_rand} are the joint and conditional probabilities for random arrangement (i.e., when $\alpha_{\zeta}^{XY} = 0$), the terms of Eq.~\ref{eq:WC} may be substituted into these relations to treat SRO effects. Consequently, the mean cohesive energy becomes:
\begin{equation}
\label{eq:ecoh_SRO}
    \bar{e} = \sum_{X} c^{X} F^{X}(\bar{\rho}^X) + \frac{1}{2} \sum_{X,Y,\zeta} c^{X}c^{Y} \left(1 -\alpha_{\zeta}^{XY}\right) Z_{\zeta} \phi_{\zeta}^{XY}  
\end{equation}
In contrast to random arrangement, where the mean electron density is single-valued, under SRO the electron density becomes a conditional mean that depends on the chemical identity of the central atom. This dependency arises because SRO differentially modifies the conditional probability of finding each neighboring species around a given central atom. The species-resolved mean electron density ($\bar{\rho}^{X}$) is described by:
\begin{equation}
\label{eq:rho_SRO}
    \bar{\rho}^{X} = \sum_{Y,\zeta} c^{Y} \left(1-\alpha_{\zeta}^{XY}\right) Z_{\zeta} \rho_{\zeta}^{Y}
\end{equation}
Eqs.~\ref{eq:ecoh_SRO} and~\ref{eq:rho_SRO} are fully determined, with $F^{X}$, $\phi_{\zeta}^{XY}$ and $\rho_{\zeta}^Y$ defined by the EAM potential, and $c$, $\alpha_{\zeta}^{XY}$ and $Z_{\zeta}$ being alloy- and structure-dependent. These relations return the equations for random arrangement at $\alpha_{\zeta}^{XY} = 0$, as expected, and reduce to a simple pair potential for $F^{X} = 0$.

With the relations for the mean cohesive energy now extended to treat SRO, we begin the analysis of the associated statistical fluctuations. We provide a brief discussion of the relations required to predict statistical fluctuations, which have been expanded in this study to include SRO effects, but refer the reader to Jagatramka et al.~\cite{jagatramka2022analytical} for the detailed derivations in random systems. Our previous work~\cite{jagatramka2022analytical} has shown how the standard deviation of cohesive energy ($\sigma_e$) may be obtained by separate treatment of the embedding ($\sigma_F$) and pair interaction ($\sigma_{\phi}$) contributions, each aggregated across chemical species and neighbor shells. To arrive at expressions for these standard deviations, we assume that the underlying distributions within and across distinct neighbor shells are uncorrelated and can therefore be directly summed. However, the covariance between the embedding and pair interaction terms is not neglected, and is specifically retained when combining terms to obtain $\sigma_e$. We adopt these assumptions \textit{a priori} and check their adequacy \textit{a posteriori} against benchmarking molecular statics (MS) calculations, noting that the assumption of statistical independence weakens as pair correlations strengthen under increasing SRO.

The statistical fluctuations in the pair interaction energy can be approached by a conditional treatment. That is, the standard deviation can be described by separately considering the distributions of pair interaction energies for each central atom. Under this consideration, the species-resolved standard deviation of the pair interaction energy ($\sigma_{\phi}^X$) is obtained from the expected value of squared deviations summed over all neighbor shells:

\begin{subequations}
\begin{align}
     \left(\sigma_{\phi}^{X} \right)^2 &= \sum_{\zeta}Z_{\zeta}\langle \left(\phi_{\zeta}^{XY} - \bar{\phi}_{\zeta}^{X}\right)^2 \rangle \label{eq:sigma_phi_species_a}\\
     &=\sum_{Y,\zeta} p_{\zeta}^{Y|X} Z_{\zeta} \left( \phi_{\zeta}^{XY} - \bar{\phi}_{\zeta}^{X} \right)^2 \label{eq:sigma_phi_species_b}\\
     &=\sum_{Y,\zeta} c^{Y}\left(1-\alpha_{\zeta}^{XY}\right) Z_{\zeta} \left( \phi_{\zeta}^{XY} - \bar{\phi}_{\zeta}^{X} \right)^2 \label{eq:sigma_phi_species_c}
\end{align}
\end{subequations}
where $\langle$ and $\rangle$ enclose the expected value operation, and $\bar{\phi}_{\zeta}^{X}$ is the mean pair interaction energy in shell $\zeta$ for a central atom $X$, with $\bar{\phi}_{\zeta}^{X} = \sum_Y c^{Y} \left(1-\alpha_{\zeta}^{XY} \right) \phi_{\zeta}^{XY}$. The species-resolved standard deviation of the pair interaction energy captures fluctuations from non-overlapping sub-populations of a larger distribution that combines all atom types. Consequently, the standard deviation of the pair interaction energy may be obtained from the law of total variance as:

\begin{equation}
\label{eq:sigma_phi}
    \sigma_{\phi}^{2} =  \overbrace{\sum_{X} c^{X} \left( \sigma_{\phi}^{X} \right)^2}^{\text{intra-species}} + \overbrace{\sum_{X<Y}c^{X}c^{Y}\left(\bar{\phi}^{X} - \bar{\phi}^{Y} \right)^2}^{\text{inter-species}}
\end{equation}
with $\bar{\phi}^X = \sum_{\zeta} \bar{\phi}_{\zeta}^X Z_{\zeta} =  \sum_{Y,\zeta}c^{Y}\left(1-\alpha_{\zeta}^{XY} \right)\phi_{\zeta}^{XY}Z_{\zeta}$. The first term in Eq.~\ref{eq:sigma_phi} represents the composition-weighted linear combination of species-resolved variances. These are conditional variances, which capture the spread in pair interaction energy from variation in neighbor-shell composition for a given central atom (i.e., the intra-species spread). Meanwhile, the second term informs the spread due to differing mean pair interaction energies across the distributions of each central atom (i.e., inter-species spread).

The fluctuations in the embedding energy may be treated using the same considerations under the law of total variance as with the pair interaction energy. That is, the embedding energy contributions to scatter can be separated by species and the total embedding energy distribution emerges from combination of these non-overlapping sub-populations. The intra-species spread, denoted here as the species-resolved standard deviation of embedding energy ($\sigma_F^{X}$), arises from a nonlinear function of the underlying distribution of electron density. In contrast to the pair interaction energy, these conditional variances do not possess an exact closed-form expression. However, our calculations have shown that the inter-species term dominates the spread in embedding energies and the intra-species term can be neglected without a significant loss of accuracy. Under this approximation, the standard deviation of the embedding energy is obtained from: 
\begin{subequations}
\label{eq:F_sigma}
\begin{align}
    \sigma_{F}^{2} &\approx  \sum_{X} c^{X} \left( \sigma_{F}^{X} \right)^2 + \sum_{X<Y}c^{X}c^{Y}\left[F^{X} \left(\bar{\rho}^{X} \right) - F^{Y} \left(\bar{\rho}^{Y} \right) \right]^2 \label{eq:F_sigma_a}\\
    &\approx \sum_{X<Y}c^{X}c^{Y}\left[F^{X} \left(\bar{\rho}^{X} \right) - F^{Y} \left(\bar{\rho}^{Y} \right) \right]^2 \label{eq:F_sigma_b}
\end{align}
\end{subequations}
Each term in Eq.~\ref{eq:F_sigma} shares the same significance as described for the standard deviation of pair interaction energy in Eq.~\ref{eq:sigma_phi}, noting that $F^X\left(\bar{\rho}^X\right)$ is single-valued and inherits the assumption that $\bar{F}^X \approx F^X\left(\bar{\rho}^X\right)$ from Varvenne et al.~\cite{varvenne2016average}.

The standard deviation of the cohesive energy may now be obtained from a combination of the embedding energy and pair interaction energy variances, while accounting for covariance between these underlying distributions:
\begin{subequations}
\label{eq:sigma_coh}
\begin{align}
    \sigma_{e}^{2} &= \sigma_{F}^{2} + \left(\frac{\sigma_{\phi}}{2}\right)^{2} + 2\,\mathrm{cov}\!\left(F, \frac{\phi}{2}\right) \label{eq:sigma_coh_a}\\
    \begin{split}
    \mathrm{cov}\!\left(F, \frac{\phi}{2}\right) &= \frac{1}{2}\sum_X c^{X} F^{X}\left(\bar{\rho}^{X}\right) \bar{\phi}^X\\ &\quad - \left[\sum_X c^{X} F^{X}\left(\bar{\rho}^{X}\right)\right] \left[\frac{1}{2}\sum_{X,Y,\zeta} c^{X} c^{Y}\left(1-\alpha_{\zeta}^{XY}\right) Z_{\zeta} \phi_{\zeta}^{XY}\right] 
    \end{split} \label{eq:covar}
\end{align}
\end{subequations}

Eqs.~\ref{eq:ecoh_SRO}--\ref{eq:sigma_coh} are complete and may be applied to crystalline CSSs of arbitrary number of components, composition, and degree of SRO. The accuracy of these relations is evaluated by comparison to MS relaxed structure calculations of the cohesive energy distributions in the FCC phases of different benchmark CSSs, as presented in Section~\ref{results_cohesive}. 

\subsection{Vacancy energy statistics in CSSs with short-range order}\label{vacancytheory}
\begin{figure}[!htbp]
    \centering
    \includegraphics[width=0.8\textwidth]{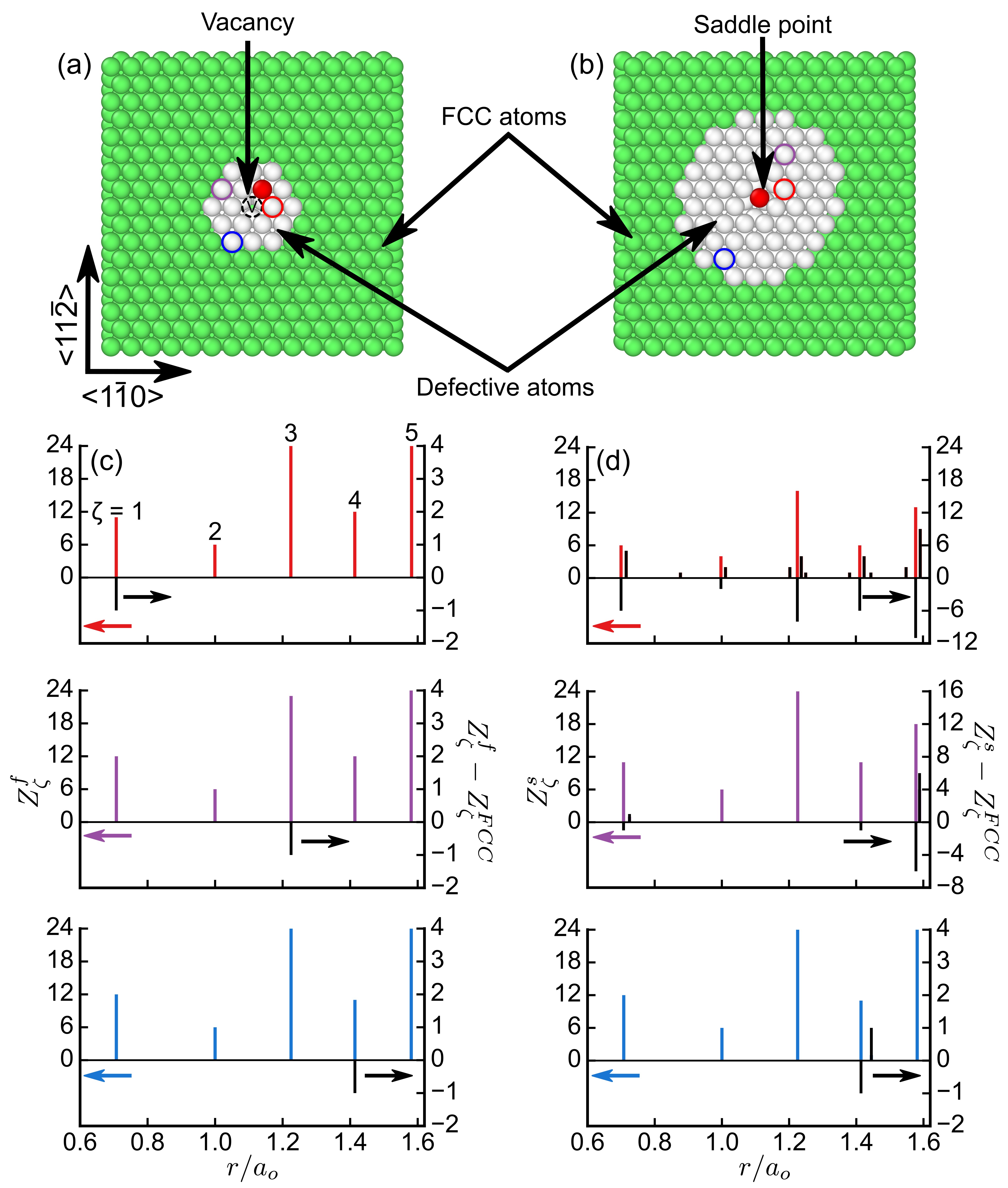}
    \caption{The FCC- and defect-coordinated sites associated with (a) vacancy formation and (b) vacancy migration in an FCC lattice viewed from a $\langle111\rangle$ zone axis. Green and white atoms denote FCC-coordinated atoms and atoms whose coordination is influenced by the defect, respectively. The red-, purple-, and blue-circled atoms represent lattice sites at progressively increasing distances from the defect (vacancy or saddle point). The excess energies of these (and other) defective sites are aggregated in the VFE and VME calculations. The structure factors for vacancy formation (c) and migration (d). The difference of these structure factors from an FCC-coordinated atom is plotted on the secondary axis in black stroke. The data are plotted up to 5 ideal FCC neighbor shells, with distances described as a ratio of the lattice parameter. The colored lines are read from the left axis and the black stroke is plotted against the right axis.}

    \label{fig:Analytical}
\end{figure}

With the equations for the cohesive energy statistics of CSSs with SRO now established, we apply these relations to lattice defects to predict the vacancy energy statistics. Our approach centers on an excess energy treatment of the vacancy formation and migration energies, where the statistics emerge from the differences between the cohesive energy distributions of a defective and reference system. This excess energy comparison can be handled directly by the relations of the mechanistic model, with analysis of the vacancy defect and base crystal differentiated by their structure factors. For the VFE, this comparison draws upon differences between the structure factors of a monovacancy and the idealized FCC lattice. Meanwhile, the VME contemplates the structure factor at the saddle point during migration against the baseline monovacancy before transit begins. Figure~\ref{fig:Analytical} overviews the shifting structure factors associated with each defect. As shown in the figure, the structure factors of atoms depend on their position relative to the defect, which results in a system of defective atoms whose aggregate excess energy contributions form the VFE and VME distributions. The significance of these excess energy contributions diminishes as distance from the defect increases. For the FCC crystal and vacancy formation process, excess energy contributions are found to become negligible beyond a radius ($r$) of $r/a_o = \sqrt{5/2}$, where $a_o$ is the FCC lattice parameter. This distance corresponds to the first five neighbor shells around a monovacancy, which coincides with the common cutoff radii for EAM potentials. For vacancy migration, the cutoff is extended up to $r/a_o \approx 3$ around the saddle point due to the misfit strain and symmetry-breaking in the lattice as the vacancy transits. Here, the structure factors for migration are comparatively denser from a loss of FCC symmetry at the vacancy saddle point. The cutoff radii are also shown schematically in Figure~\ref{fig:Analytical}, where the green atoms beyond the cutoff are taken as FCC coordinated and do not contribute to excess energy. Atoms within the cutoff radius are considered to be in a defective condition and are shown in white. The atomic configurations shown here are visualized using the Open Visualization Tool (OVITO)~\cite{stukowski2009visualization}. The structure factors for the FCC crystal are calculated directly from the ideal lattice geometry. However, the monovacancy and saddle-point configurations are extracted from MS and climbing-image nudged elastic band (CI-NEB) calculations~\cite{Henkelman2000}, with all cutoff radii determined by sensitivity analysis. Section~\ref{MS_calc} provides further details on these methods. 

The mean vacancy formation energy ($\bar{E}_{VFE}$) can be written as the sum of excess energies arising from the differences between defective and FCC mean cohesive energies for all sites within the vacancy-influenced cutoff:
\begin{equation}
\label{eq:VFE_mean}
\bar{E}_{VFE} =
\sum_{\zeta=1}^{\zeta_{cut}^f}
Z_{\zeta}^f\left(\bar{e}_{\zeta}^{f} - \bar{e}_{FCC}\right)
\end{equation}
where $Z_{\zeta}^{f}$ is the number of sites in neighbor shell $\zeta$ from the monovacancy, $\zeta_{cut}^f$ is the cutoff neighbor shell for vacancy formation, $\bar{e}_{\zeta}^{f}$ is the mean cohesive energy of an atom in neighbor shell $\zeta$ from the vacancy, and $\bar{e}_{FCC}$ is the mean cohesive energy of an FCC-coordinated atom. Both $\bar{e}_{\zeta}^{f}$ and $\bar{e}_{FCC}$ are obtained using Eq.~\ref{eq:ecoh_SRO} with the appropriate structure factor. The structure factors used in the calculation of each site-specific $\bar{e}$ are distinct and differ from those referring to the vacancy defect (see Figure~\ref{fig:Analytical}). The mean vacancy migration energy ($\bar{E}_{VME}$) is expressed as:
\begin{equation}
\label{eq:VME_mean}
\bar{E}_{VME} =\sum_{\zeta=1}^{\zeta_{cut}^{s}}Z_{\zeta}^{s}(\bar{e}_{\zeta}^{s} - \bar{e}_{FCC})  - \bar{E}_{VFE}
\end{equation}
where $\zeta_{cut}^s$ is the cutoff neighbor shell for vacancy migration at the saddle point and $\bar{e}_{\zeta}^{s}$ is the mean cohesive energy of an atom in neighbor shell $\zeta$ from the migrating atom at the saddle point. $\bar{e}_{\zeta}^{s}$ is again obtained by Eq.~\ref{eq:ecoh_SRO} by way of the corresponding structure factors. Here, the summation term represents the excess cohesive energy of the saddle-point configuration relative to the FCC crystal. The subtraction of $\bar{E}_{VFE}$ accounts for excess energy associated with the prerequisite process of vacancy formation prior to migration. Collectively, Eqs.~\ref{eq:VFE_mean} and~\ref{eq:VME_mean} demonstrate how mean vacancy energies may be written as the aggregate excess energies defined by the structure factors of the defective and reference crystal environment. A complete derivation of the mean vacancy formation and migration expressions is provided in~\ref{AA}.

The standard deviation of the vacancy formation and migration energies may be contemplated using the same excess energy approach. That is, the spreads of excess energies at each site that interacts with the vacancy defect are individually treated and then aggregated through statistical combination of variances. We begin with the site excess energies of vacancy formation. Within the cutoff radius, each site has an excess energy of $e_{f} - e_{FCC}$, representing the difference of the cohesive energy at that location for an atom in the monovacancy and FCC system, respectively. The individual standard deviations at each site ($\sigma_{\Delta f}$) may be obtained following the rules for the subtraction of statistical distributions from:
\begin{subequations}
\begin{align}
\sigma_{\Delta f}^{2} &= \langle \left[\left(e_{f} - e_{FCC}\right) - \left(\bar{e}_{f}-\bar{e}_{FCC}\right)\right]^{2}\rangle\\
&= \sigma_{f}^2 + \sigma_{FCC}^2 - 2\sigma_{f}\sigma_{FCC} \label{eq:VFE_std_site}
\end{align}
\end{subequations}
where $\sigma_{f}$ and $\sigma_{FCC}$ are the standard deviations of the cohesive energies for the same site in the monovacancy and FCC crystal systems, respectively. These standard deviations can be calculated using the relations from Section~\ref{Theory} and are differentiated by their structure factors. The terms are assumed to be perfectly correlated, which allows for the covariance (i.e., $\mathrm{cov}(e_{f},e_{FCC})$) to be simplified to $\sigma_{f}\sigma_{FCC}$. This assumption is supported by a covariance analysis that is provided in the Supplementary Materials. To estimate the spread in the vacancy formation energies, $\sigma_{\Delta f}$ must be aggregated across all sites within the cutoff radius. Here, we assume a perfect correlation between excess energies in the same neighbor shell around the monovacancy, but otherwise no correlation between neighbor shells. Consequently, the standard deviation of the vacancy formation energy ($\sigma_{VFE}$) may be obtained from:
\begin{equation}
\label{eq:VFE_std}
\sigma_{VFE}^2 = \sum_{\zeta=1}^{\zeta_{cut}^f}\left(Z_\zeta^{f}\sigma_{\Delta f_\zeta}\right)^2
\end{equation}
where $\sigma_{\Delta f_\zeta}$ is the standard deviation of the site excess energy in neighbor shell $\zeta$. This term is the same as calculated by Eq.~\ref{eq:VFE_std_site} but has been aggregated with other sites in the same neighbor shell by the structure factor $Z_{\zeta}^{f}$. Using a similar approach, the standard deviation of the vacancy migration energy ($\sigma_{VME}$) is given as:
\begin{equation}
\label{eq:VME_std}
\sigma_{VME}^2 = \sum_{\zeta=1}^{\zeta_{cut}^s}\left(Z_\zeta^s\sigma_{\Delta s_\zeta}\right)^2 + \sigma_{VFE}^2
\end{equation}
where $\sigma_{\Delta s_\zeta}$ is analogous to $\sigma_{\Delta f_\zeta}$ but refers to excess energies in the neighbor shells of the saddle-point configuration. A complete derivation of $\sigma_{VFE}$ and $\sigma_{VME}$ is provided in~\ref{AB}.

\subsection{Effects of fluctuations on vacancy-mediated properties} \label{Effective}
With the relations for the statistical parameters of the vacancy energy distributions complete, we contemplate the implications of fluctuations in defect energies on vacancy-mediated properties. For this purpose, we examine the vacancy concentration and vacancy diffusivity in CSSs, which are traditionally described by fixed values of defect energies. The equilibrium vacancy concentration ($C_{v}$) is calculated from classical thermodynamic relations as~\cite{Mehrer2007}:
\begin{equation}
\label{eq:C_v}
C_v = \exp{\left\{\frac{S_f}{k}\right\}}\exp{\left\{\frac{-E_{VFE}}{kT}\right\}}
\end{equation}
where $S_f$, $k$ and $T$ are the vacancy formation entropy, Boltzmann constant, and temperature, respectively. Here, the effects of pressure and volume change are routinely neglected such that the enthalpy of formation can be approximated by $E_{VFE}$, and $S_f$ is taken as $2k$, following the work of Burton~\cite{burton1972vacancy}. The vacancy diffusivity ($D_{v}$) is classically given by the following relationship~\cite{Mehrer2007}:
\begin{equation}
\label{eq:D_v}
    D_v =  \frac{1}{6}f_v d^{2} Z_1\nu \exp{\left\{\frac{-E_{VME}}{kT}\right\}}
\end{equation}
$Z_1$ is the first-shell vacancy coordination, which is 12 for an FCC material. $f_v$ is the correlation factor, $d$ is the jump distance (i.e., $a_o/\sqrt{2}$), and $\nu$ is the attempt frequency, which is typically taken as $\approx 10^{12}$~s$^{-1}$ for metals and alloys including CSSs~\cite{du2022chemical}.

In evaluation of Eqs.~\ref{eq:C_v} and~\ref{eq:D_v}, the vacancy formation and migration energies are traditionally taken as constants. Here, we develop relations to explore the effect of statistical scatter on the values of equilibrium vacancy concentration and vacancy diffusivity. These predictions are later validated through comparison to direct simulations and the available literature data. For this purpose, we pursue an effective energy barrier approach that replaces the singular values of $E_{VFE}$ and $E_{VME}$ with barriers weighted by the statistics of the underlying vacancy formation and migration energy distributions. To achieve this outcome, we evaluate the expected value of the Boltzmann factor of the relevant energy barrier. From examination of Eqs.~\ref{eq:C_v} and~\ref{eq:D_v}, this analysis is in the form $\exp{\left\{-aE\right\}}$, where $a$ is a constant and $E$ is drawn from the distribution of vacancy formation or migration energies. Our previous work has shown that this form of expected value can be evaluated by integrating the Gaussian probability density function over all possible energy barriers~\cite{jagatramka2023evolution}:
\begin{subequations}
\label{eq:gaussian_average}
\begin{align}
\begin{split}
\langle\exp{\left\{-aE\right\}}\rangle
&= \frac{1}{\sigma\sqrt{2\pi}} \Bigg[\int_{0}^{\infty} \exp\left\{-aE\right\}
\exp\left\{-\frac{1}{2}\left(\frac{E-\bar{E}} {\sigma}\right)^{2}\right\}dE\\
&\quad + \int_{-\infty}^{0} \exp\left\{-\frac{1}{2}
   \left(\frac{E-\bar{E}}{\sigma}\right)^{2}\right\}dE \Bigg]
\end{split}\\[6pt]
\begin{split}
&= \exp\left\{-a\left(\bar{E}
   - \frac{a\sigma^{2}}{2}\right)\right\}
\left(\frac{1}{2}
   + \frac{1}{2}\operatorname{erf}\left(
     \frac{\bar{E} - a\sigma^{2}}
          {\sqrt{2}\sigma}\right)\right) \\
&\quad + \left(\frac{1}{2}
   - \frac{1}{2}\operatorname{erf}\left(
     \frac{\bar{E}}
          {\sqrt{2}\sigma}\right)\right)
\end{split}
\label{eq:gaussian_average_closed}
\end{align}
\end{subequations}
where $a = 1/kT$, and $\bar{E}$ and $\sigma$ are the mean and standard deviation of the distribution, respectively. For the equilibrium vacancy concentration, these parameters are drawn from the vacancy formation energy distribution, whereas calculation of the expected value for the vacancy diffusivity proceeds from the vacancy migration energy statistics. The first integral in Eq.~\ref{eq:gaussian_average} reflects integration over all possible positive energy barriers. The second integral treats the small but non-zero probability of finding barriers with $E < 0$. This artifact is a natural consequence of the tails of Gaussian statistics and leads to instability in the analysis. Consequently, this term fixes the Boltzmann factor of these barriers at $\exp{\left\{-aE\right\}} = 1$ for all $E \leq 0$, effectively setting a hard limit of $E = 0$ on the energy barrier. We now determine the effective energy barrier ($\hat{E}$) by equating $\exp{\left\{-a\hat{E}\right\}} = \langle\exp{\left\{-aE\right\}}\rangle$, which yields:

\begin{align}
\begin{split}
\hat{E}
&= -\frac{1}{a} \ln \Bigg[
   \exp\left\{-a\left(\bar{E}
   - \frac{a\sigma^{2}}{2}\right)\right\}\left(\frac{1}{2}
   + \frac{1}{2}\operatorname{erf}\left(
     \frac{\bar{E} - a\sigma^{2}}
          {\sqrt{2}\sigma}\right)\right) \\
&\quad + \left(\frac{1}{2}
   - \frac{1}{2}\operatorname{erf}\left(
     \frac{\bar{E}}
          {\sqrt{2}\sigma}\right)\right)
   \Bigg]
\end{split}
\label{eq:expected}
\end{align}
This relation is similar to one proposed for fluctuating VFEs by Ruban~\cite{Ruban2016}. However, for Gaussian distributions, that formulation admits negative energies at the low-energy tail, an effect that becomes significant for broad distributions. An examination of the extremes of Eq.~\ref{eq:expected} shows that $\hat{E}$ approaches $\bar{E}$ from below as the spread vanishes (i.e., $\sigma \rightarrow 0$). Furthermore, $\hat{E} \rightarrow kT\ln\left(2\right)$ in the infinite fluctuation limit (i.e., as $\sigma \rightarrow \infty$). A key interpretation from these limiting cases is that $\hat{E}$ is less than $\bar{E}$, with the separation growing at increasing spread and saturating at $\bar{E} - kT\ln\left(2\right)$. In the following sections, we explore the effects and sensitivity of the vacancy-mediated transport properties to these fluctuations.

\section{Computational methodology} \label{methodology}
The relations of the mechanistic model are validated using two different types of MS simulations: MS relaxation calculations and CI-NEB simulations. The former is used to obtain the statistical distributions of cohesive energies and VFEs by comparing the relaxed energies of CSSs with and without a monovacancy. Meanwhile, the latter provides the VME statistics by comparing the relaxed saddle-point energy of vacancy transport with a monovacancy baseline. These simulation procedures are well established in the community~\cite{23, 34, 35, wang2018energetics, del2011formation, guo2024properties}. Furthermore, we separately develop kinetic Monte Carlo (kMC) simulations to validate the predictions for vacancy diffusivity under fluctuating VME barriers. The details of these simulations are provided in the Supplementary Materials. 

\subsection{Selection of CSSs} \label{materials}
To explore the mechanistic model and effect of fluctuations on vacancy-mediated transport properties, we have selected specific compositions within the FeNiCr and FeNiCrCoCu systems as our benchmarks. Our rationale for selecting these systems is multifold. First, FeNiCr has technological relevance as the base system of 304-type austenitic stainless steels (i.e., Fe\(_{73}\)Ni\(_8\)Cr\(_{19}\) by mole), for which some vacancy-mediated transport property data are available in the literature. Meanwhile, the quinary FeNiCrCoCu system enables sequential examination of CSSs with increasing chemical complexity, whereby we have selected ternary, quaternary, and quinary sub-systems from this alloy family. In addition to these qualities, EAM-style interatomic potentials are readily available to drive simulations for both systems~\cite{deluigi2021simulations,RD3}. For brevity, we restrict our vacancy studies to Fe\(_{73}\)Ni\(_8\)Cr\(_{19}\) from the FeNiCr system. Within the FeNiCrCoCu system, we select the equimolar NiCrCo, NiCrCoCu, and FeNiCrCoCu CSSs. However, we validate the mechanistic model relations for cohesive energy statistics across a much broader composition space in both systems. Different levels of SRO are considered in both the cohesive energy and the vacancy studies. Although we acknowledge that these EAM potentials may not have been specifically fit to model interactions across such a broad alloy range, we emphasize that our mechanistic model draws upon their coefficients and does not introduce any new adjustable parameters or pathways for error in capturing interatomic interactions.

\subsection{Atomistic simulations} \label{MS_calc}
To execute CSS simulations, single crystals are constructed by a computer script using the idealized FCC structure factors. In this script, atoms are randomly selected and assigned elements in proportion to the target alloy composition, which produces randomly arranged CSSs. To produce CSSs with varying SRO, we leverage a Monte Carlo-based atom swapping algorithm. Here, pairs of unlike atoms are randomly selected and their positions are exchanged. This exchange is accepted when the resulting WC parameters of the swapped system reduce the residual relative to the WC parameter targets, which are constrained by the sum rule, $\sum_{Y}c^{Y}\alpha_{\zeta}^{XY} = 0$. The Monte Carlo swapping loop terminates once the residual drops below a prescribed threshold. This approach differs from other swapping algorithms reported in the SRO literature as it does not rely on a Metropolis-style acceptance criterion~\cite{Wang2022}. However, this method is appropriate for our interests, which are to explore a range of SRO configurations, rather than identify the thermodynamically stable SRO. To limit the configurational search space, we restrict exploration of SRO effects to the first nearest-neighbor shell and uniformly vary all heteroatomic pair correlations. For example, in the FeNiCr system the heteroatomic pair correlations are identically set as $\alpha^{XY}_{1} =\alpha^{\mathrm{FeNi}}_{1}=\alpha^{\mathrm{FeCr}}_{1}=\alpha^{\mathrm{NiCr}}_{1}$. An analogous WC parameter selection is enforced in the FeNiCrCoCu system. Consequently, the homoatomic pair correlations, $\alpha_{1}^{XX}$, are fixed by chemical normalization to $\alpha_{1}^{XX} = -\alpha_{1}^{XY}\left(1-c^{X}\right)/c^{X}$. Furthermore, we track the changes in higher neighbor shell WC parameters (i.e., $\alpha_{\zeta > 1}^{XY}$) that arise from adjustment of $\alpha_{1}^{XY}$, and propagate these variations into the mechanistic model calculations. Further details on the Monte Carlo swap algorithm and some example data are provided in the Supplementary Materials.

All atomistic simulations are performed using the large-scale atomic/molecular massively parallel simulator (LAMMPS)~\cite{plimpton1995fast, Thompson2022}. Interatomic interactions in the FeNiCr and FeNiCrCoCu systems are modeled using the EAM potentials from Bonny et al.~\cite{RD3} and Deluigi et al.~\cite{deluigi2021simulations}, respectively. The EAM cutoff radii are set to 0.56~nm for FeNiCr and 0.58~nm for the FeNiCrCoCu alloys, as specified by the respective potentials. The simulation cell axes are aligned to the $\langle 1\bar{1}0 \rangle$, $\langle 11\bar{2} \rangle$, and $\langle 111 \rangle$ crystallographic directions, with periodic boundary conditions enforced along all axes. All simulations measure at least \(7 \times 7 \times 7\)~nm\(^3\) and contain more than 40,000 atoms. The system size was validated by sensitivity analysis to ensure convergence of key energies and the elimination of finite-size effects and is far beyond the size-effect thresholds identified in previous work~\cite{manzoor2021machine}.

The cohesive energies of CSSs are determined directly from MS relaxation simulations of perfect FCC crystals using conjugate gradient energy minimization. The simulation cell boundaries are also allowed to change dimension to further relax the structure during energy minimization. The FCC lattice parameters, which are needed to set structure factors, are determined from the second peak of the radial distribution function of the relaxed structures. From this analysis, lattice parameters of 0.3510, 0.3531, 0.3546, and 0.3549~nm are measured for the Fe\(_{73}\)Ni\(_8\)Cr\(_{19}\) and equimolar NiCrCo, NiCrCoCu, and FeNiCrCoCu systems, respectively. The per-atom energies are exported from the MS simulations, from which the means and standard deviations of the cohesive energy distributions are directly determined.

\begin{figure}[!t]
    \centering
    \includegraphics[width=\textwidth]{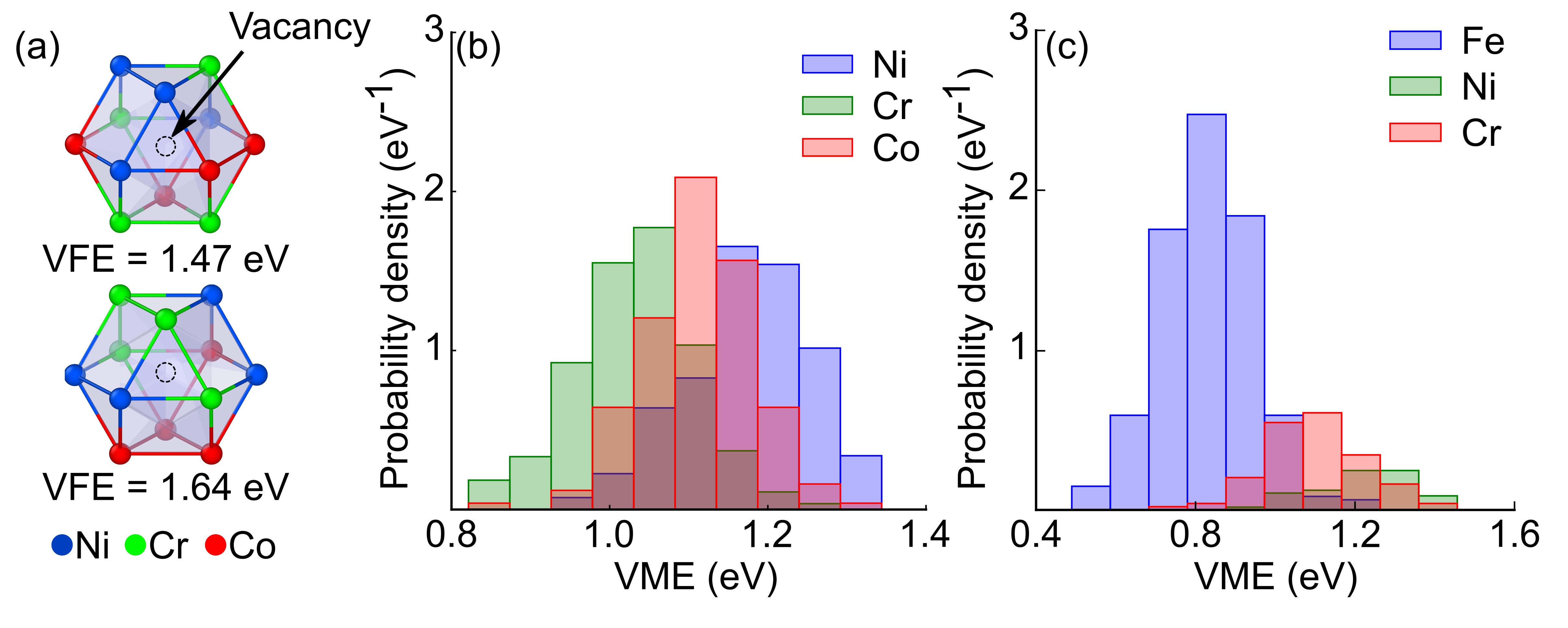}
    \caption{(a) Nearest-neighbor arrangements around two examples of vacancies in the equimolar NiCrCo system. Although the nearest-neighbor chemistries are identical, rearrangement of the neighboring atoms dramatically changes the VFE, contributing to the spread in defect energies. Probability density histograms of the VMEs calculated from CI-NEB simulations for the (b) equimolar NiCrCo and (c) Fe\(_{73}\)Ni\(_8\)Cr\(_{19}\) alloys. The results are resolved by species of the transported atom, which helps visualize the contributions of chemistry and configuration to the spread in energies.}
    \label{fig:methods}
\end{figure}

To obtain the VFE distribution, we generate a simulation cell and conduct MS relaxation using the methods described above. An atom is then randomly deleted from the relaxed structure and the system is subjected to a second MS relaxation step. Upon convergence of energy minimization, the system energy is recorded. The VFE is calculated using the relation $E_{VFE} = E_{f} - E_{FCC} + \mu$, where $E_{f}$ is the total energy of the relaxed monovacancy system, $E_{FCC}$ is the total energy of the relaxed pristine system, and $\mu$ is taken as the energy of the removed atom in the pristine relaxed structure to align with the model treatment (see~\ref{AB}). We replicate this simulation at least 450 times and then take the mean and standard deviation of the collected $E_{VFE}$. A replica sensitivity analysis is provided in the Supplementary Materials, which shows convergence of the statistical parameters at around 200 replicas. Two representative cases corresponding to VFEs for the randomly arranged equimolar NiCrCo system are shown in Figure~\ref{fig:methods}a. As shown in the figure, although the nearest-neighbor chemistries are identical, the difference in VFEs is 0.17~eV, highlighting the sensitivity of defect energies to the local atomic arrangement. 

The VME distribution is obtained from CI-NEB simulations of nearest-neighbor vacancy transport in relaxed monovacancy systems produced using the methods described above. In each simulation, a nearest neighbor of the monovacancy is randomly selected and then migrated into the vacant site along a $\langle 110\rangle$ direction to complete vacancy transport. The \texttt{fix neb} method in LAMMPS with the QuickMin minimization scheme is used to execute migration simulations and converge saddle-point configurations. All VMEs are reported as forward barriers and are therefore determined as the difference between the saddle-point energy and the relaxed energy of the initial monovacancy system. Each CI-NEB simulation consists of 16 images that are initially equally spaced along the transport pathway, and all migration calculations are replicated at least 450 times. A replica sensitivity analysis showing convergence of the VME distribution is provided in the Supplementary Materials. Examples of the VME distributions for the randomly arranged equimolar NiCrCo and Fe\(_{73}\)Ni\(_8\)Cr\(_{19}\) alloys are provided in Figure~\ref{fig:methods}b and~\ref{fig:methods}c, respectively. Here, the VME data are differentiated by the species of the transported atom, illustrating the chemical and configuration-related contributions to the spread in migration energies. We note the distinct intra-species and inter-species behaviors that contribute to the shape of the VME distribution. In equimolar NiCrCo, the mean energies and spreads are similar among the constituents. Consequently, the overall distribution is well approximated by a unimodal Gaussian, with its shape dominated by intra-species effects. In contrast, the Fe\(_{73}\)Ni\(_8\)Cr\(_{19}\) distribution exhibits a much larger inter-species contribution that arises from the separation in mean energies. Combined with the larger Fe content, these features positively skew the distribution. We discuss the implications of skewed distributions on the accuracy of the mechanistic model in subsequent sections.

Atomistic simulations are also leveraged to determine the individual structure factors for the atoms within the cutoff radii of the vacancy defect, as discussed in Section~\ref{vacancytheory}. For this purpose, we simulate the relaxed structures of the monovacancy and saddle-point migration configurations using the methods described above. To generalize these measurements, we select pure FCC iron as our benchmark for these vacancy defect structure factors under the EAM potential of Bonny et al.~\cite{RD3}. Distances in the structure factors are normalized by $a_o$ and then scaled to each CSS system using the corresponding lattice parameter. Examples of structure factors for atoms within the defect radius cutoff of the vacancy defects are provided in Figure~\ref{fig:Analytical}. This step obviates the need for exhaustive relaxation simulations for each CSS of interest and streamlines high-throughput calculations with the mechanistic model.

\section{Results and discussion}
\subsection{Validation of cohesive energy relations for CSSs with short-range order} \label{results_cohesive}
This section presents the results of MS calculations of cohesive energies for a range of CSSs with both random and SRO arrangements. These results are compared against the predictions from the relations of the mechanistic model. The range of alloy compositions is investigated along three distinct enrichment paths for each system. For FeNiCr, these are Fe$_x$Ni$_{(1-x)/2}$Cr$_{(1-x)/2}$, Fe$_{(1-x)/2}$Ni$_x$Cr$_{(1-x)/2}$, and Fe$_{(1-x)/2}$Ni$_{(1-x)/2}$Cr$_x$, corresponding to Fe-enriched, Ni-enriched, and Cr-enriched compositions, respectively. For FeNiCrCoCu, the enrichment paths are $\mathrm{Fe}_{x}(\mathrm{NiCrCo})_{(1-x)/2}\mathrm{Cu}_{(1-x)/2}$, $\mathrm{Fe}_{(1-x)/2}(\mathrm{NiCrCo})_{x}\mathrm{Cu}_{(1-x)/2}$, and $\mathrm{Fe}_{(1-x)/2}(\mathrm{NiCrCo})_{(1-x)/2}\mathrm{Cu}_{x}$, representing Fe-enriched, equimolar NiCrCo-enriched, and Cu-enriched compositions, respectively. Although we acknowledge that the FCC phase is not stable for many of these compositions, we nonetheless explore these structures as a theoretical exercise. The range of compositions investigated along these paths is fixed by the normalizations of the WC parameter such that some compositions are forbidden at specific levels of SRO. The natural range of $p_{\zeta}^{Y|X}$ combined with the WC parameter sum rule, $\sum_{Y}c^{Y}\alpha_{\zeta}^{XY} = 0$, and the restrictions we have set on heteroatomic pair correlations (i.e., identical $\alpha_{\zeta}^{XY}$) impose a lower bound on the permissible compositions of $c^{Y} \geq -\alpha_{\zeta}^{XY}/\left(1-\alpha_{\zeta}^{XY}\right)$. All positive values of $\alpha_{\zeta}^{XY}$  are permissible. However, composition thresholds emerge for negative values of $\alpha_{\zeta}^{XY}$, with minimum concentrations rising as the heteroatomic WC parameter decreases.

\begin{figure}[!t]
\centering
\includegraphics[width=1.0\textwidth]{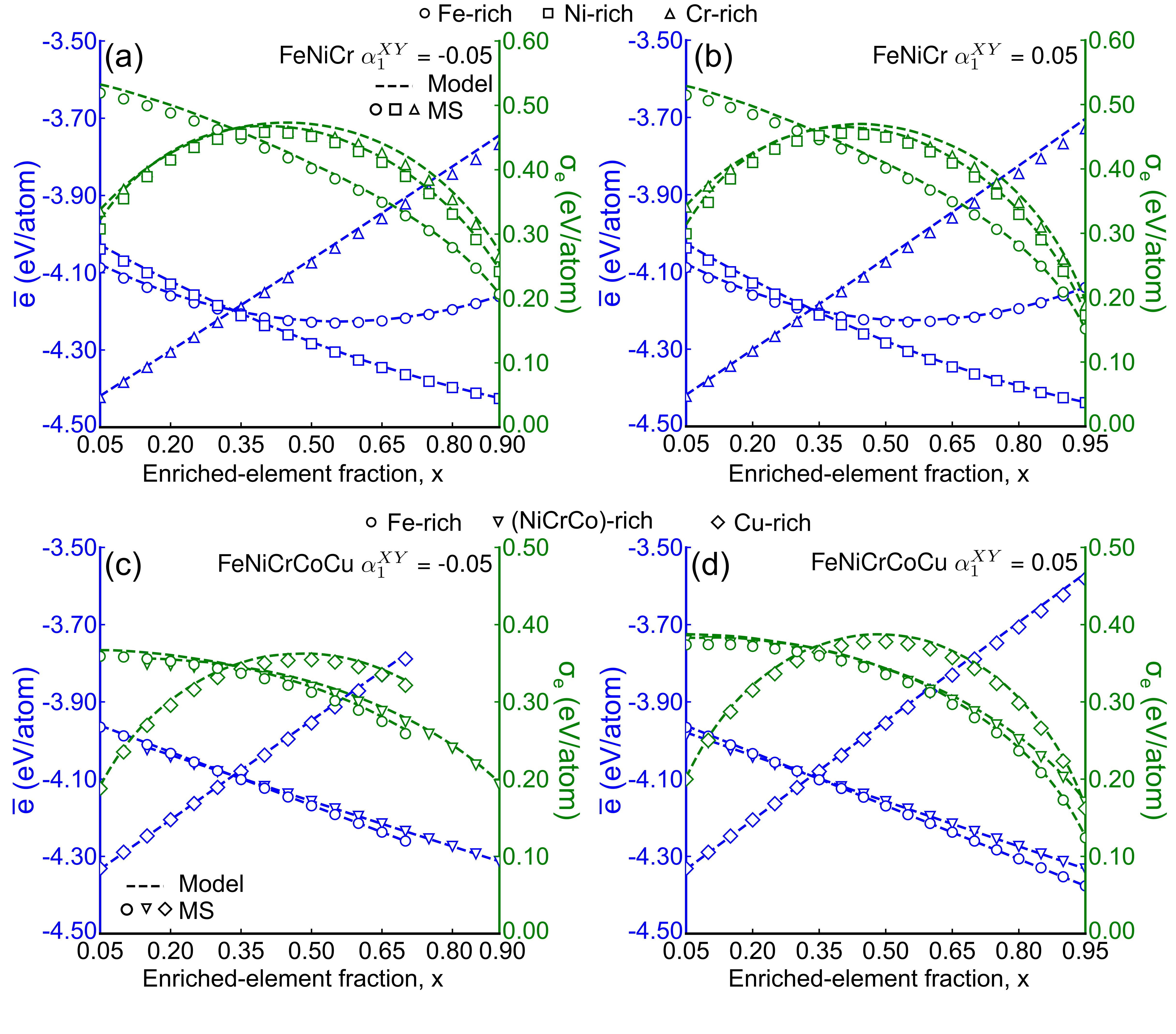}
\caption{Comparison of the mechanistic model predictions against MS relaxed structure calculations for the mean cohesive energy (blue) and its spread (green). The comparison is mapped over a range of permissible alloy compositions at different levels of SRO. Results are provided for the FeNiCr system for (a) $\alpha_{1}^{XY} = -0.05$ and (b) $0.05$, evaluated along the Fe-rich (Fe$_x$Ni$_{(1-x)/2}$Cr$_{(1-x)/2}$), Ni-rich (Fe$_{(1-x)/2}$Ni$_x$Cr$_{(1-x)/2}$), and Cr-rich (Fe$_{(1-x)/2}$Ni$_{(1-x)/2}$Cr$_x$) composition paths. (c, d) Analogous results are presented for the FeNiCrCoCu system for Fe-rich (Fe$_x$(NiCrCo)$_{(1-x)/2}$Cu$_{(1-x)/2}$), NiCrCo-rich (Fe$_{(1-x)/2}$(NiCrCo)$_x$Cu$_{(1-x)/2}$), and Cu-rich (Fe$_{(1-x)/2}$(NiCrCo)$_{(1-x)/2}$Cu$_x$) alloys.}
\label{fig:SRO}
\end{figure}

Figure~\ref{fig:SRO} provides the cohesive energy statistics for compositional sweeps across the selected alloys of the FeNiCr and FeNiCrCoCu systems at $\alpha_{1}^{XY}$ values of $-0.05$ and $0.05$. We have selected these WC parameters for this compositional study, as they represent the termini of observed SRO for the few quantified measurements available in the experimental literature~\cite{He2024}. As shown in the figure, the mechanistic model captures the ground-truth MS calculations with excellent accuracy for mean cohesive energies and their standard deviations. Note that MS calculations include lattice distortion effects that are not captured by FCC structure factors, but these appear to have a negligible effect on cohesive energy statistics. Furthermore, the strong agreement observed here supports the assumptions undertaken in model development. For FeNiCr, the selected alloys exhibit a multitude of patterns across sweeps, including monotonic and non-monotonic behaviors of varying compositional sensitivities and asymmetries in the compositional dependencies of cohesive energy parameters, all of which are quantitatively captured by the mechanistic model. An equally close agreement is mirrored in the studies of the FeNiCrCoCu system, indicating that our approach retains its accuracy in alloys with increasing configurational complexity.

Examination of the trends in the standard deviations of cohesive energies provides interesting insights into configurational entropy effects. For instance, in FeNiCr, the Ni- and Cr-rich paths reach peak standard deviations near, but not at, the equimolar composition ($x \approx 0.4$), suggesting that maximizing configurational entropy contributes to the spread. However, the spread is maximized near the binary NiCr limit along the Fe-rich path. Similarly, the equimolar FeNiCrCoCu alloy (read in Figure~\ref{fig:SRO} as Fe\(_{20}\)(NiCrCo)\(_{60}\)Cu\(_{20}\)) exhibits a lower standard deviation relative to several other compositions. From these results, we note the general observation that while configurational entropy may have some modest effect on the spread in the atomic-scale properties of CSSs, any correlation is likely to be weak. The community currently has high interest in the computational design of medium and high entropy alloys with large spreads in defect energies. These results emphasize that non-equimolar compositions and reduced component systems cannot be immediately excluded from design considerations.

\begin{figure}[!b]
\centering
\includegraphics[width=1.0\textwidth]{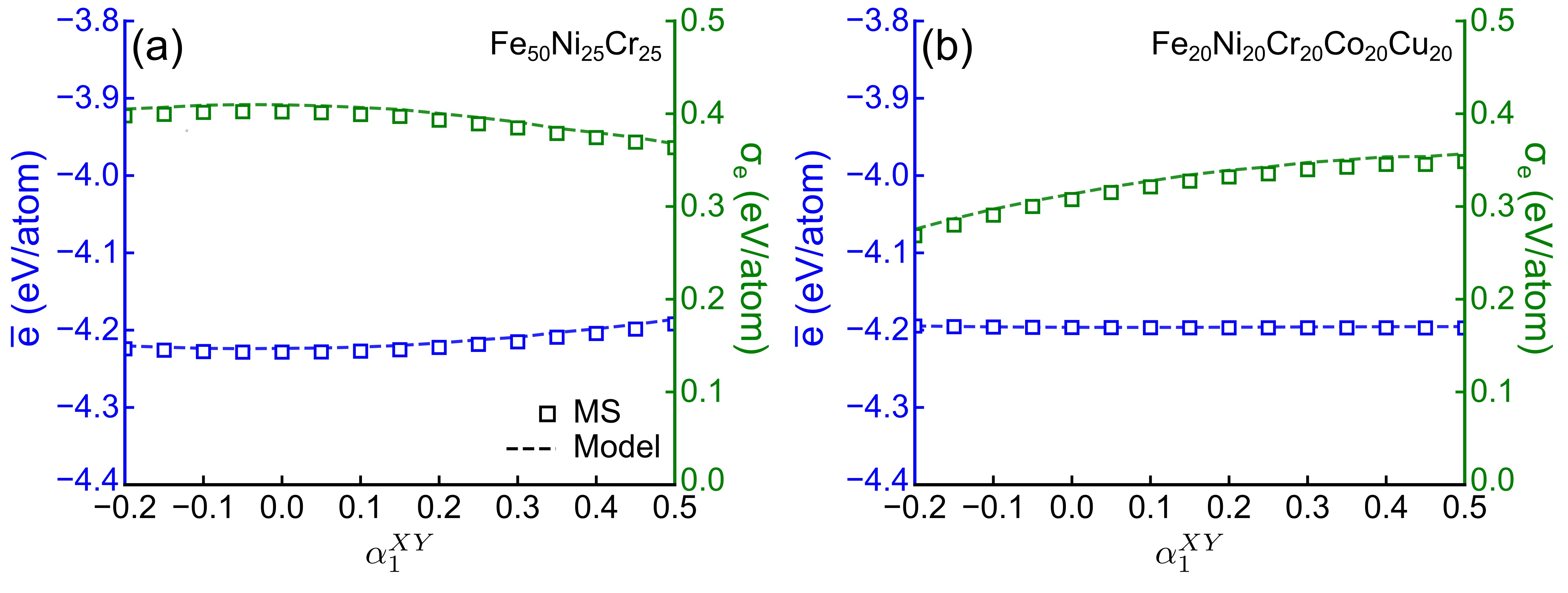}
\caption{Mean cohesive energy and standard deviation mapped over a range of the WC parameter ($\alpha_{1}^{XY}$), for (a) $\mathrm{Fe_{50}Ni_{25}Cr_{25}}$ and (b) equimolar FeNiCrCoCu. The comparison demonstrates that the mechanistic model accurately captures SRO-induced variations in both the mean value and the standard deviation of the cohesive energy over a wide range of $\alpha_{1}^{XY}$.}
\label{fig:SRO_alpha}
\end{figure}

Figure~\ref{fig:SRO_alpha} provides the cohesive energy statistics for fixed compositions while sweeping broadly across the SRO configurational space. Although such a large range of SRO has not been measured experimentally, these extremes of the WC parameter do appear in several computational studies (e.g., Ref.~\cite{Sheriff2024}). We again note the excellent agreement between the mechanistic model and MS calculations. Compared to the compositional sweep provided in Figure~\ref{fig:SRO}, SRO exhibits a weaker coupling to cohesive energy statistics. However, inspection of the results does provide interesting observations. For instance, the mean cohesive energy shows a negligible sensitivity to SRO in the quinary system. Meanwhile, the system exhibits a monotonic increase in the spread of cohesive energies as nearest-neighbor heteroatomic interactions are increasingly suppressed (i.e., at positive $\alpha_{1}^{XY}$) and clustering becomes dominant. Collectively, the results of Figures~\ref{fig:SRO} and~\ref{fig:SRO_alpha} highlight the flexibility of the mechanistic model to capture chemical and structural effects in CSSs of arbitrary composition and complexity.

\begin{figure}[!t]
\centering
\includegraphics[width=1.0\textwidth]{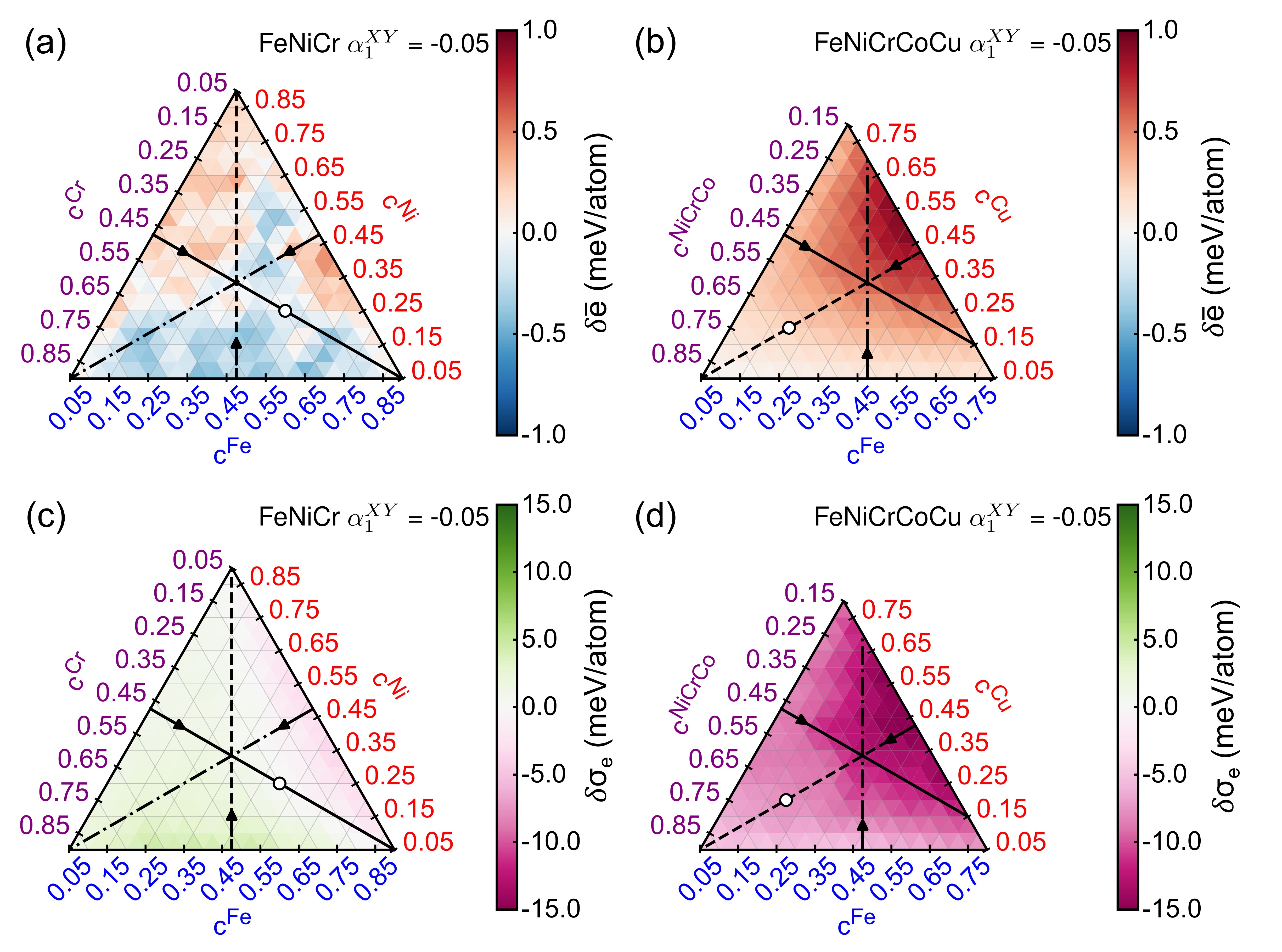}
\caption{The influence of SRO on the cohesive energy statistics. This comparison is presented as deviations in statistical parameters for CSSs with $\alpha_{1}^{XY} = -0.05$ referenced against random arrangement. (a,b) Variation in the mean cohesive energy for FeNiCr and FeNiCrCoCu, respectively. (c,d) The corresponding variation in the distribution spreads. The maps reveal a weak coupling of SRO to the mean cohesive energy in these systems. However, a substantially larger effect is observed in the standard deviation, with the magnitude and sign of the variation depending on alloy composition. The lines in each diagram plot the enrichment paths explored in Figure~\ref{fig:SRO} and the markers indicate the compositions studied in Figure~\ref{fig:SRO_alpha}.}
\label{fig:SRO_difference}
\end{figure}

To further study the significance of SRO in these systems, Figure~\ref{fig:SRO_difference} presents model predictions for the variations in the mean cohesive energy ($\delta\bar{e}$) and standard deviation ($\delta\sigma_{e}$) mapped across the alloy composition space. The model predictions are presented as deviations in statistical parameters for $\alpha_{1}^{XY} = -0.05$, which are referenced against the values for random arrangement ($\alpha_{\zeta}^{XY} = 0$). The corresponding variation diagrams for the clustering configuration with $\alpha_{1}^{XY} = 0.05$ are provided in the Supplementary Materials and display qualitatively similar patterns. For both FeNiCr and FeNiCrCoCu, the mean cohesive energy exhibits a weak dependence on SRO across all compositions, demonstrating that the first statistical moment is comparatively insensitive to chemical ordering. In contrast, the spread in energies displays a markedly stronger, albeit still mild, composition dependence with both the magnitude and the sign of $\delta\sigma_{e}$ varying. One interpretation is that SRO primarily modifies the distribution and dispersion of cohesive energies, rather than their arithmetic mean. Although we cannot generalize this metallurgy beyond the systems studied here, this observation reinforces the need to explicitly account for SRO effects when predicting vacancy-mediated properties in CSSs. 

This compositional mapping also provides a practical highlight of our mechanistic model. That is, the alloy configurational space (i.e., both chemistry and SRO) can be searched almost instantaneously. The computational time required to solve the relations of the mechanistic model is trivial using modest desktop computing hardware ($\sim$1~s in all systems studied), whereas measurement by direct simulation over the entire configurational space requires between hours and weeks to complete, depending on system complexity. We have written scripts to facilitate model calculations, which can be freely accessed through a GitHub repository~\cite{AdvancedMaterialsandMicrostructuresLabattheUniversityofIllinoisChicago}.

\subsection{Statistics of the vacancy formation and migration energy distributions} \label{distribution}
\begin{figure}[!t]
    \centering
    \includegraphics[width=0.9\textwidth]{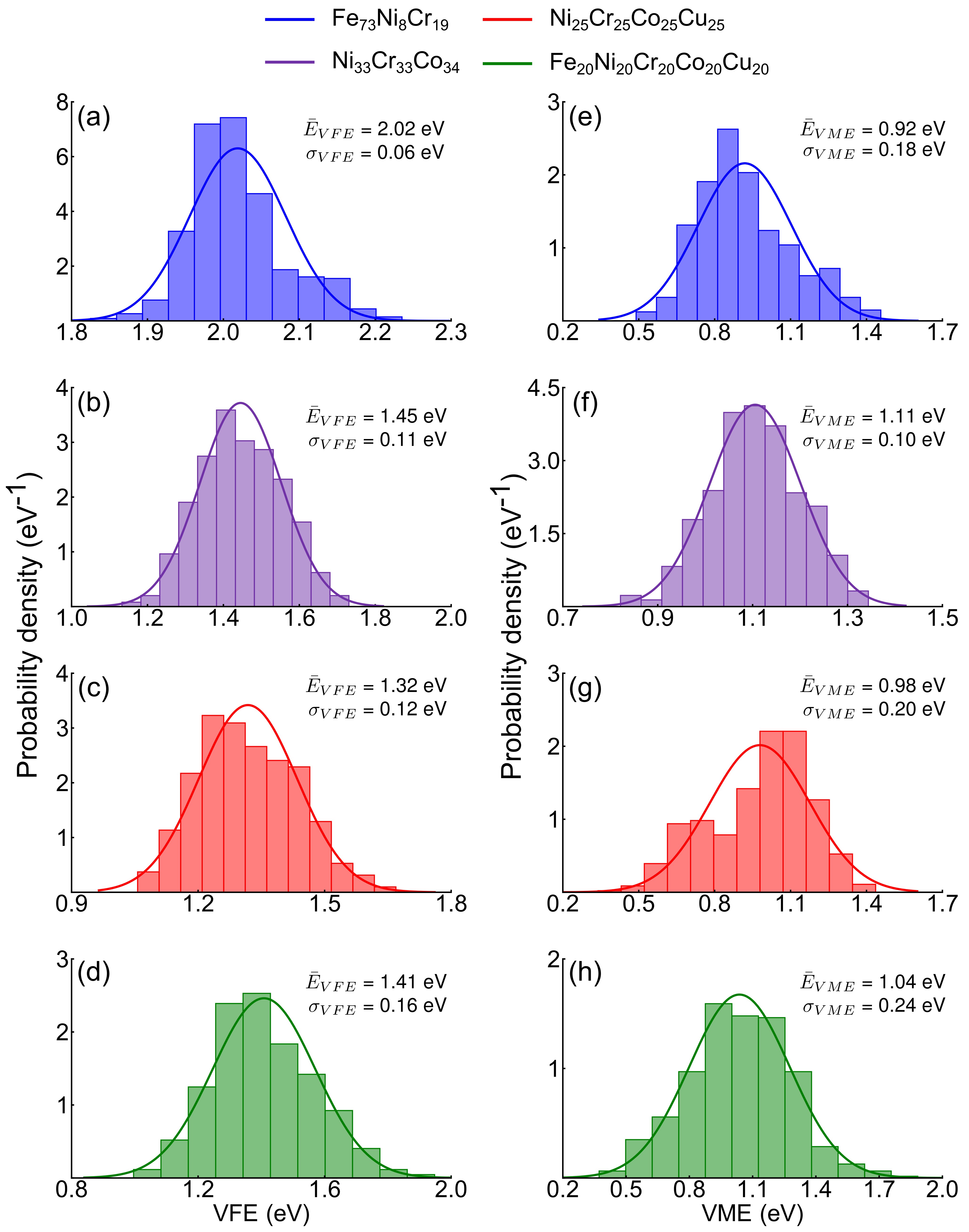}
    \caption{Probability density histograms of the (a)-(d) VFEs and (e)-(h) VMEs measured by atomistic simulations for randomly arranged CSSs ($\alpha_{\zeta}^{XY} = 0$). A unimodal Gaussian has been fit to each defect energy and is overlaid on the histogram data. The means and standard deviations of the fitted Gaussians are provided, which serve as our validation benchmarks for the mechanistic model.}
    \label{fig:PDF}
\end{figure}

Figure~\ref{fig:PDF} presents VFEs and VMEs measured from the MS and CI-NEB simulations for Fe\(_{73}\)Ni\(_8\)Cr\(_{19}\) and the equimolar NiCrCo, NiCrCoCu, and FeNiCrCoCu CSSs under random arrangement. The data are presented as probability density histograms, which have been fit with a unimodal Gaussian to good approximation in most cases. The mean and standard deviation of the fitted distributions for each CSS are provided in the figure. These data have also been summarized in Table~\ref{tab:defect_energies}. We measure mean VFEs of 2.02 and 1.41~eV, for Fe\(_{73}\)Ni\(_8\)Cr\(_{19}\) and equimolar FeNiCrCoCu, respectively. These values are in excellent agreement with literature reports of 2.01--2.04~eV~\cite{27} and 1.42~eV~\cite{deluigi2021simulations}, respectively, which are derived using atomistic calculations of similar compositions. The narrowest VFE distribution is measured from Fe\(_{73}\)Ni\(_8\)Cr\(_{19}\) with $\sigma_{VFE} = 0.06$~eV. In contrast, the other ternary system examined within this group (equimolar NiCrCo) exhibits a broader VFE distribution with $\sigma_{VFE} = 0.11$~eV. The quaternary and quinary systems exhibit further broadening in the spread of VFEs, reaching up to $\sigma_{VFE} = 0.16$~eV. We hesitate to attribute this increase in spread to a rise in configurational entropy, given the findings of the previous section.

The trends in VME distributions exhibit patterns different from those observed for formation energies. For instance, Fe\(_{73}\)Ni\(_{8}\)Cr\(_{19}\) exhibits a larger spread in its VMEs relative to equimolar NiCrCo. Furthermore, although the distribution can be fairly approximated by a unimodal Gaussian, the underlying histogram appears positively skewed. This skew arises from differences in the mean energies of the underlying species-resolved migration distributions, as discussed in Section~\ref{MS_calc}. In contrast to Fe\(_{73}\)Ni\(_{8}\)Cr\(_{19}\), equimolar NiCrCo appears to be well captured by the unimodal Gaussian. However, upon introduction of Cu in the quaternary system, the skew re-emerges, albeit with a negative character. The quinary system exhibits the largest spread in migration energies, but appears symmetric upon the addition of Fe. Furthermore, it is interesting to note that skewness appears to be most evident in the VMEs, and can emerge from alloys that present nearly symmetric VFE distributions (e.g., equimolar NiCrCoCu). This observation suggests that skew is not intrinsic to either defect type, but arises when constituents interact differently with the local defect structure, which separates the species-resolved sub-distributions, as in the VME distributions shown in Figure~\ref{fig:methods}. We have also explored the effects of SRO on the VFE and VME in Fe\(_{73}\)Ni\(_{8}\)Cr\(_{19}\) and equimolar FeNiCrCoCu. The results of Gaussian fitting against this set of simulations are provided in Table~\ref{tab:defect_energies} and the probability density histograms are plotted in the Supplementary Materials. Here, we note a minor change ($\sim$10s of meV) in the VFE and VME statistical parameters under modest SRO, with the variations marginally larger in the quinary system. This behavior tracks with the underlying trends in the cohesive energy reported in the previous section. That is, chemistry appears to be the dominant contributor to changes in VFE and VME, while SRO plays a modest but non-negligible role in defining vacancy energy distributions. We explore the sensitivities of vacancy-mediated properties to changes in the statistical parameters of defect energy distributions in Section~\ref{Effective_Results} to further support these observations.

\begin{table}[!t]
    \scriptsize
    \centering
    \begin{threeparttable}
    \caption{Vacancy energy statistics as calculated by the mechanistic model and measured by atomistic simulations. All energies are reported in eV.}
    \label{tab:defect_energies}
    \begin{tabular}{lc cccc cccc}
        \toprule
        & & \multicolumn{4}{c}{\textbf{VFE}} & \multicolumn{4}{c}{\textbf{VME}} \\
        \cmidrule(lr){3-6} \cmidrule(lr){7-10}
        & & \multicolumn{2}{c}{$\bar{E}_{VFE}$} & \multicolumn{2}{c}{$\sigma_{VFE}$}
          & \multicolumn{2}{c}{$\bar{E}_{VME}$} & \multicolumn{2}{c}{$\sigma_{VME}$} \\
        \cmidrule(lr){3-4} \cmidrule(lr){5-6} \cmidrule(lr){7-8} \cmidrule(lr){9-10}
        \textbf{CSS} & $\alpha^{XY}_{1}$
        & Model & MS & Model & MS
        & Model & CI-NEB & Model & CI-NEB \\
        \midrule
        Fe$_{73}$Ni$_{8}$Cr$_{19}$                       & 0     & 2.02 & 2.02 & 0.03 & 0.06 & 0.95 & 0.92 & 0.05 & 0.18 \\
        Ni$_{33}$Cr$_{33}$Co$_{34}$                      & 0     & 1.45 & 1.45 & 0.20 & 0.11 & 1.11 & 1.11 & 0.21 & 0.10 \\
        Ni$_{25}$Cr$_{25}$Co$_{25}$Cu$_{25}$             & 0     & 1.32 & 1.32 & 0.18 & 0.12 & 0.98 & 0.98 & 0.22 & 0.20 \\
        Fe$_{20}$Ni$_{20}$Cr$_{20}$Co$_{20}$Cu$_{20}$    & 0     & 1.42 & 1.41 & 0.22 & 0.16 & 1.04 & 1.04 & 0.26 & 0.24 \\
        Fe$_{73}$Ni$_{8}$Cr$_{19}$                       & $-0.05$ & 2.02 & 2.02 & 0.03 & 0.06 & 0.95 & 0.92 & 0.05 & 0.18 \\
        Fe$_{73}$Ni$_{8}$Cr$_{19}$                       & $0.05$ & 2.02 & 2.02 & 0.03 & 0.07 & 0.94 & 0.92 & 0.06 & 0.19 \\
        Fe$_{20}$Ni$_{20}$Cr$_{20}$Co$_{20}$Cu$_{20}$    & $-0.05$ & 1.41 & 1.39 & 0.19 & 0.15 & 1.04 & 1.03 & 0.23 & 0.22 \\
        Fe$_{20}$Ni$_{20}$Cr$_{20}$Co$_{20}$Cu$_{20}$    & $0.05$ & 1.44 & 1.43 & 0.25 & 0.18 & 1.04 & 1.04 & 0.29 & 0.25 \\
        \bottomrule
    \end{tabular}
    \end{threeparttable}
\end{table}

\begin{figure}[!t]
    \centering
    \includegraphics[width=1\textwidth]{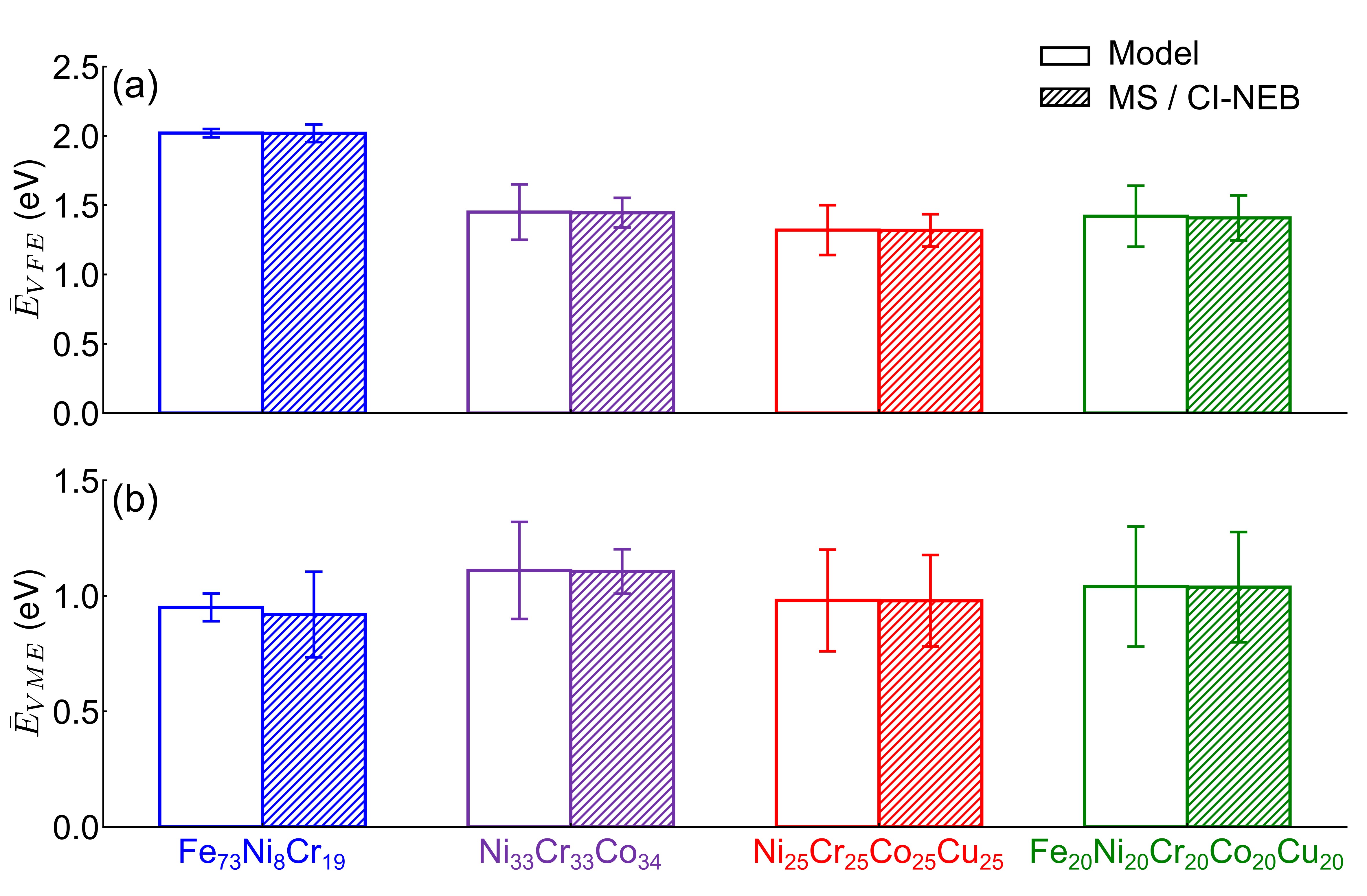}
    \caption{A comparison of the measured and predicted statistical parameters for the (a) VFE and (b) VME distributions for randomly arranged CSSs ($\alpha_{\zeta}^{XY} = 0$). Error bars denote one standard deviation.}
    \label{fig:Rand_Histogram}
\end{figure}

\begin{figure}[!t]
    \centering
    \includegraphics[width=1\textwidth]{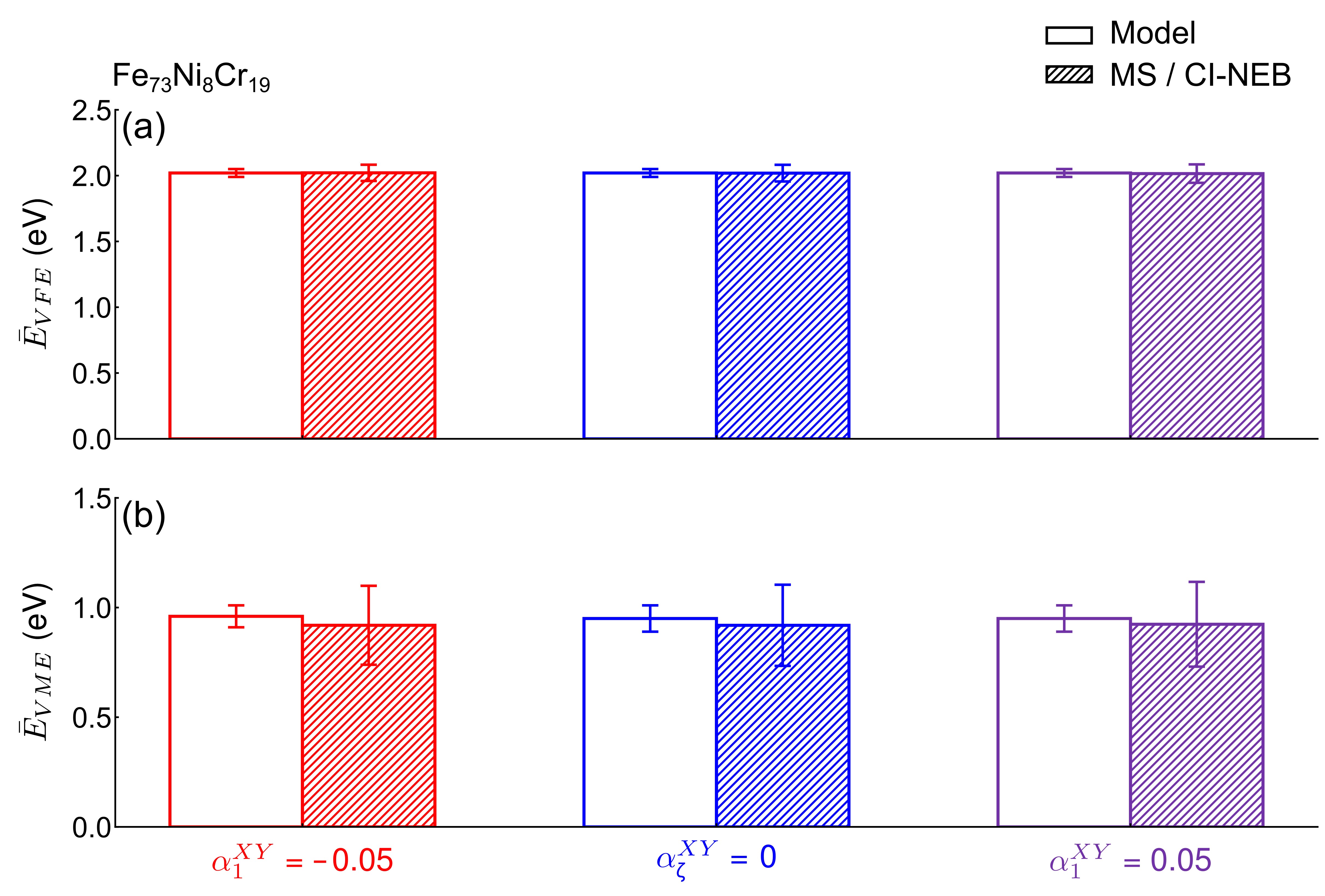}
    \caption{The effect of SRO on the vacancy energies of the Fe$_{73}$Ni$_{8}$Cr$_{19}$ alloy. The statistical parameters measured from atomistic simulations are compared against the predictions of the mechanistic model for (a) the VFE and (b) the VME. Error bars denote one standard deviation.}
    \label{fig:SRO_Histogram}
\end{figure}

\begin{figure}[!t]
    \centering
    \includegraphics[width=1\textwidth]{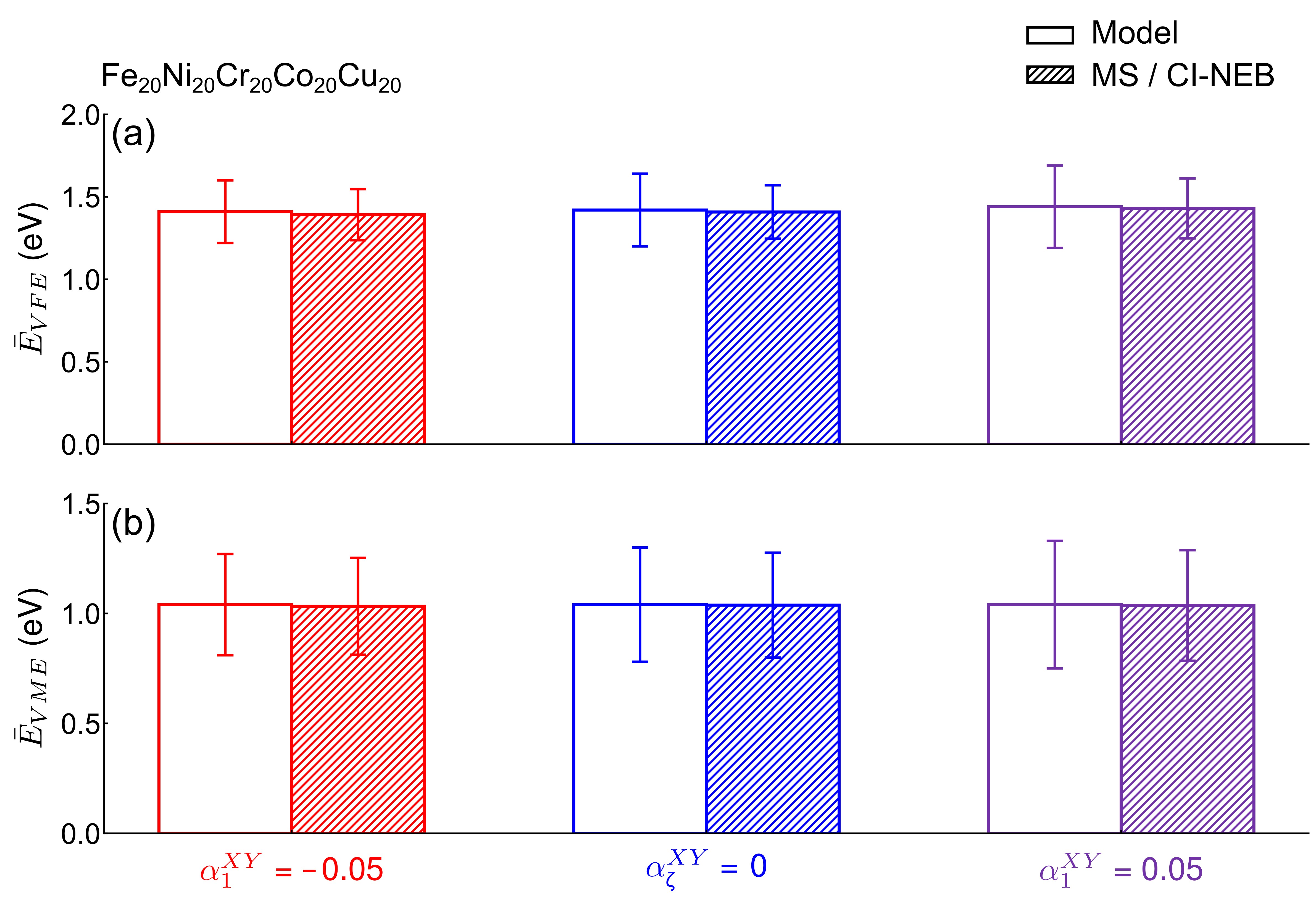}
    \caption{The effect of SRO on the vacancy energies of equimolar FeNiCrCoCu. The statistical parameters measured from atomistic simulations are compared against the predictions of the mechanistic model for (a) the VFE and (b) the VME. Error bars denote one standard deviation.}
    \label{fig:SRO_Histogram_quinary}
\end{figure}

We now leverage the atomistic measurements of the vacancy energy distributions in these CSS systems to validate the mechanistic model relations developed in Section~\ref{vacancytheory}. Figure~\ref{fig:Rand_Histogram} compares the model predictions with the measured statistical parameters of the VFEs and VMEs for randomly arranged CSSs, while Figure~\ref{fig:SRO_Histogram} and Figure~\ref{fig:SRO_Histogram_quinary} present the same information for Fe\(_{73}\)Ni\(_{8}\)Cr\(_{19}\) and equimolar FeNiCrCoCu under different states of SRO, respectively. The model predictions are also listed in Table~\ref{tab:defect_energies}. As shown in the figures and table, the mean vacancy energies are predicted with excellent accuracy. We also note that the model captures the essential trends across the different CSSs and also predicts a weak effect of SRO on the vacancy energy distributions. For mean VFEs, the deviations from direct atomistic measurements are at most 0.02~eV, translating to an error of $\sim$1\%. For mean VMEs, the error is marginally larger, with deviations reaching up to 0.03~eV (i.e., less than 4\%) for Fe\(_{73}\)Ni\(_{8}\)Cr\(_{19}\). We note that this level of error is within the range of literature discrepancies for direct simulations and is significantly lower than the mean absolute errors of $\sim$1~eV reported in a recent machine-learning-based study~\cite{choudhary2023can}. Relative to these methods, we emphasize that the mechanistic model offers comparable or greater accuracy at significantly higher throughput. It does not require execution of atomistic simulations or large pre-computed training datasets, possesses no adjustable parameters outside of those embedded in the underlying interatomic potentials, and may be executed using modest desktop hardware. 

The model predictions for the spread in the VFEs and VMEs reproduce the collected atomistic data with good accuracy. However, relative to the mean vacancy energies, the error in the predictions of standard deviations is larger. Here, the deviations from atomistic measurements range from 0.03--0.09~eV for VFEs and 0.01--0.13~eV for the VMEs. The spreads in vacancy energy distributions of CSSs are not commonly reported in the literature, due in large part to a lack of high-throughput approaches, and it is therefore difficult to find a benchmark for model performance. Nonetheless, the errors for the alloys studied remain small fractions of the distribution means. The largest error occurs in the Fe\(_{73}\)Ni\(_{8}\)Cr\(_{19}\) system, where the model systematically underpredicts the VME distribution spread. The model does not resolve the identity of the migrating atom. We therefore attribute this behavior to the missing inter-species component of the spread, which is large in this alloy (see Figure~\ref{fig:methods}c) and also produces the skew. Beyond this system, we foresee scenarios where large inter-species differences in vacancy energies or constituent dilution further increase skew, leading to separation of distributions at the extremes. Such systems represent a significant departure from the metallurgy of CSSs and we therefore expect model accuracy to significantly degrade in these cases.

\subsection{Vacancy-mediated properties under fluctuating energy barriers} \label{Effective_Results}
\begin{figure}[!b]
    \centering
    \includegraphics[width=0.9\textwidth]{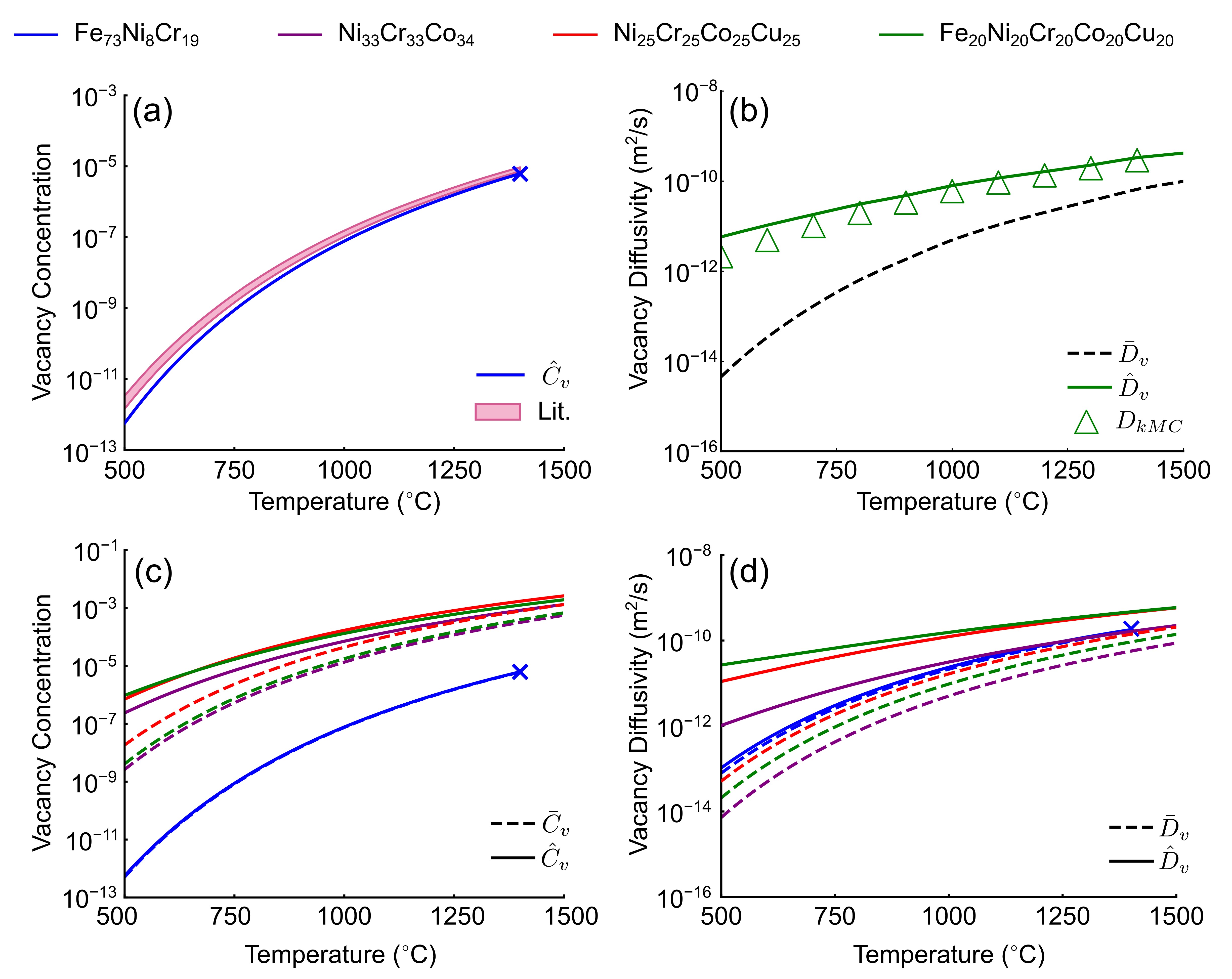}
    \caption{Temperature dependence of vacancy-mediated properties in randomly arranged CSSs as predicted by the mechanistic model. (a) Equilibrium vacancy concentration as a function of temperature for Fe$_{73}$Ni$_8$Cr$_{19}$, as predicted by Eq.~\ref{eq:C_v} using the effective barrier $\hat{E}_{VFE}$ given by Eq.~\ref{eq:expected}. These predictions are compared against concentrations calculated by Eq.~\ref{eq:C_v} using effective formation energies digitized from Zhang et al.~\cite{27}, which were obtained from DFT calculations in different supercells of a randomly arranged FeNiCr alloy of similar composition. The shaded region spans the results from the different supercells. (b) Vacancy diffusivity as a function of temperature for Fe$_{20}$Ni$_{20}$Cr$_{20}$Co$_{20}$Cu$_{20}$ calculated by using Eq.~\ref{eq:D_v}. The plot compares calculations for diffusivity ($\bar{D}_v$ and $\hat{D}_v$, respectively) using the average ($\bar{E}_{VME}$) and effective barriers ($\hat{E}_{VME}$, Eq.~\ref{eq:expected}) against kMC vacancy transport simulations ($D_{kMC}$). The temperature dependence of the (c) equilibrium vacancy concentration and (d) vacancy diffusivity for the four CSSs as predicted by average and effective barriers. Property predictions for Fe$_{73}$Ni$_8$Cr$_{19}$ have been truncated at the solidus temperature of 304 stainless steel ($\sim$1400$~^\circ$C), which is marked by a blue x.}
    \label{fig:Prop}
\end{figure}

In this section, we explore the effects of fluctuations in the formation and migration energies on vacancy-mediated properties. For this purpose, we compare the predictions of the equilibrium vacancy concentration and vacancy diffusivity under mean and effective barriers using the relations of Section~\ref{Effective}. By ignoring statistical spread, the former calculation disregards the kinetics of lower-energy barriers available at vacancy energy distribution tails. While this treatment is often sufficient for pure materials and dilute alloys, we explore its deficiencies in CSSs with comparatively broad vacancy energy spectra through effective barrier property calculations. The mean barrier equilibrium vacancy concentration $\left(\bar{C}_v\right)$ and vacancy diffusivity $\left(\bar{D}_v\right)$ are calculated from Eqs.~\ref{eq:C_v} and~\ref{eq:D_v} with $\bar{E}_{VFE}$ and $\bar{E}_{VME}$, respectively. For the effective barrier treatment ($\hat{C}_v$ and $\hat{D}_v$, respectively), we perform the same calculation but use $\hat{E}_{VFE}$ and $\hat{E}_{VME}$ as determined from Eq.~\ref{eq:expected} for each CSS distribution. Figure~\ref{fig:Prop} presents the vacancy-mediated property results for randomly arranged CSSs over a wide range of temperatures. To validate calculations, we have compared vacancy concentration predictions against literature data and vacancy diffusivity predictions against kMC simulations. 

Figure~\ref{fig:Prop}a plots the equilibrium vacancy concentration for Fe$_{73}$Ni$_{8}$Cr$_{19}$ against data obtained from Zhang et al.~\cite{27} for an FeNiCr alloy of similar composition. The literature range is calculated using effective formation energies obtained from DFT-based monovacancy simulations in different supercells. As shown in the figure, our effective energy calculation closely matches the literature data at high temperatures and falls slightly below the literature range as temperature decreases. This difference arises mainly from the broader VFE distributions reported by Zhang et al., which reflect differences in the treatment of interatomic interactions (i.e., DFT versus EAM potential) and the limited sampling of small supercells. An additional contributor is the treatment of the chemical potential (see~\ref{AB}). Figure~\ref{fig:Prop}c provides both the mean and effective barrier predictions for all the CSSs examined in this study under random arrangement. We note that in all cases the effective barrier model predicts a higher equilibrium vacancy concentration. This follows the anticipated thermodynamics emerging from consideration of statistical distribution tails and is consistent with effective barrier calculations, where $\hat{E}_{VFE} < \bar{E}_{VFE}$. Phrased differently, this behavior is a consequence of comparatively rare low-energy configurations that disproportionately contribute to barrier-limited processes, causing mean barrier models to systematically underestimate vacancy-mediated properties. We also note that the degree of separation between mean and effective barrier predictions scales with the spreads in the underlying statistical distributions. For instance, at $\sigma_{VFE} = 0.03$~eV, the predictions for the equilibrium vacancy concentration nearly overlap for Fe$_{73}$Ni$_{8}$Cr$_{19}$. However, the ratio of effective to mean barrier predictions nears $3 \times 10^2$ at 500~$^\circ$C for the equimolar FeNiCrCoCu alloy, where the largest value of $\sigma_{VFE}$ is measured. At low temperatures, the predictions for the equimolar CSSs exhibit interesting behavior. In contrast to the mean barrier vacancy concentrations, which decay significantly below $\sim$800~$^\circ$C, the effective barrier predictions remain high. This behavior arises from the low-energy tail of the VFE distribution, whose contribution grows as temperature decreases (i.e., the $-a\sigma^2/2$ dependence of $\hat{E}$).

Figure~\ref{fig:Prop}b compares the results of the vacancy diffusivity predictions for the mean and effective barrier treatments of the randomly arranged equimolar FeNiCrCoCu alloy. These predictions are plotted against kMC simulations of vacancy transport that are created using inputs from CI-NEB simulations. For both the mean and effective barrier treatments, the correlation factor $f_v$ is obtained from the corresponding kMC simulation. The details of kMC simulations are provided in the Supplementary Materials. As shown in the figure, vacancy diffusivities predicted using the effective barrier exhibit excellent agreement with the kMC, whereas predictions based on the mean barrier remain systematically lower across the entire temperature range investigated. A modest discrepancy between $\hat{D}_v$ and $D_{kMC}$ emerges at low temperatures. These discrepancies arise largely from differences in the distribution spreads between the mechanistic model and CI-NEB simulations, which increase in importance at lower temperatures where thermal activation is suppressed. Figure~\ref{fig:Prop}d provides the expanded diffusivity dataset for all CSSs considered in this study under random arrangement. For comparison purposes in this subfigure, all vacancy diffusivities are calculated as random walks (i.e., $f_v = 1$). Here, we note similar trends to those observed with the equilibrium vacancy concentration. That is, the separation between the mean and effective barrier property treatments scales monotonically with the spread in the underlying energy barrier distribution, with Fe$_{73}$Ni$_{8}$Cr$_{19}$ and equimolar FeNiCrCoCu appearing at the lower and upper extremes, respectively. We also note the same patterns in the temperature dependencies across vacancy-mediated property predictions. That is, for both the equilibrium vacancy concentration and vacancy diffusivity the mean and effective barrier predictions converge at high temperatures and exhibit their largest separation at the lowest temperatures. Once more, this behavior emerges due to the amplified significance of Boltzmann statistics in thermally activated processes at lower temperatures. Phrased differently, rare low-energy configurations contribute disproportionately at lower temperatures where thermal activation is suppressed. Conversely, differences in energy barriers become less significant at higher temperatures where sufficient thermal energy is available to access kinetically unfavorable configurations. The effect reported here is similar to trends in the vacancy transport in CSSs noted by Xing et al.~\cite{xing2024vacancy}.

\begin{figure}[!b]
    \centering
    \includegraphics[width=\textwidth]{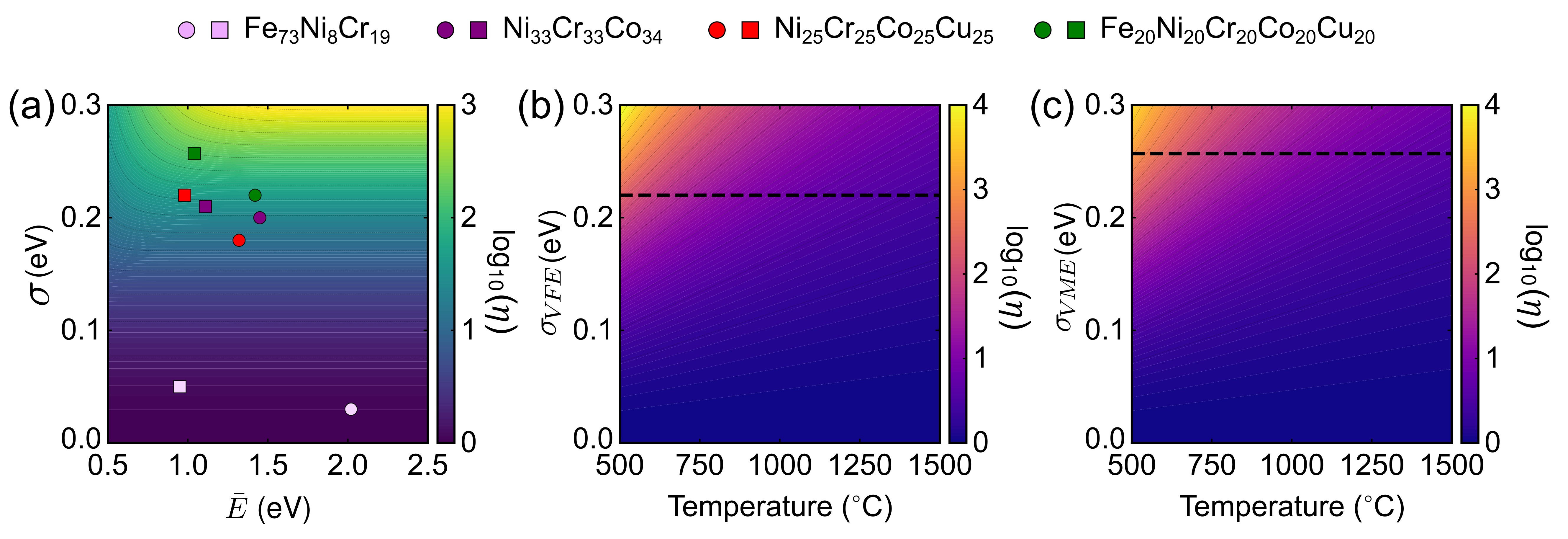}
    \caption{The sensitivity of vacancy-mediated properties to the statistical parameters of vacancy energy distributions. (a) The enhancement factor, $\eta$, plotted as a function of the mean vacancy energy and its standard deviation at $650~^\circ\text{C}$. The statistical parameters of the vacancy energies examined in this study are plotted as circles and squares for the VFE and VME distributions, respectively. Temperature dependence of the (b) vacancy concentration and (c) vacancy diffusivity enhancement factors as a function of $\sigma$ for the randomly arranged equimolar FeNiCrCoCu CSS ($\bar{E}_{VFE} = 1.42$~eV and $\bar{E}_{VME} = 1.04$~eV). The dashed horizontal lines in (b) and (c) are the model predicted distribution spreads of $\sigma_{VFE} = 0.22$~eV and $\sigma_{VME} = 0.26$~eV, respectively.}
    \label{fig:StatContours}
\end{figure}

To further explore the impact of vacancy energy fluctuations beyond the specific alloy systems investigated, Figure~\ref{fig:StatContours} presents contour plots of the relative property change over a wide range of statistical parameters. For this purpose, we define the enhancement factor ($\eta$) as the ratio of the effective barrier to mean barrier values of the vacancy-mediated property such that the pre-exponential factors for both equilibrium vacancy concentration and diffusivity cancel. Under these considerations, we define the enhancement factor as:
\begin{equation}
\label{eq:enhance}
    \eta \equiv \left(\frac{\hat{C}_v}{\bar{C}_v}\ \mathrm{or}\ \frac{\hat{D}_v}{\bar{D}_v}\right) = \exp\left\{\frac{\bar{E} - \hat{E}}{kT}\right\}
\end{equation}
where $\bar{E}$ and $\hat{E}$ are the mean and effective barriers, respectively, as obtained from the vacancy formation or migration energy distributions. Figure~\ref{fig:StatContours}a presents the enhancement factors at $650~^\circ\text{C}$ over a range of mean barriers and spreads. This intermediate temperature is selected as it represents a typical operating temperature for next-generation advanced nuclear reactor designs~\cite{Klueh2007}, where vacancy-mediated diffusional flow is anticipated to be an important deformation mechanism. The statistical parameters of the VFE and VME distributions based on the mechanistic model predictions are plotted as circles and squares, respectively. As shown in the figure, the enhancement factor reaches $10^3$ near the extremes mapped, with most CSSs falling between $\sim$10 and $\sim$$10^2$. Furthermore, the enhancement contours appear insensitive to the value of the mean barrier under most combinations of statistical parameters. However, this behavior changes as the spread increases or the mean barrier moves towards zero. This behavior is a consequence of the zero-energy lower bound placed on the effective barriers considered in this study, with the sensitivity to $\bar{E}$ rising as temperatures decrease. Figures~\ref{fig:StatContours}b and c illustrate the temperature dependence of the enhancement factor of randomly arranged equimolar FeNiCrCoCu for the equilibrium vacancy concentration and vacancy diffusivity, respectively. Here, the contour maps are calculated using the mean vacancy energies, $\bar{E}_{VFE} = 1.42$~eV and $\bar{E}_{VME} = 1.04$~eV, respectively, while the standard deviations vary. As shown in the figure, the enhancement factors are largest at lower temperatures, reaching $10^4$. At higher temperatures, thermal activation increases and the significance of the separation between mean and effective energy barriers diminishes. Collectively, these results demonstrate that vacancy-mediated properties can be varied by several orders of magnitude through small changes to the spreads in the underlying vacancy energy distributions.

\begin{figure}[!b]
    \centering
    \includegraphics[width=\textwidth]{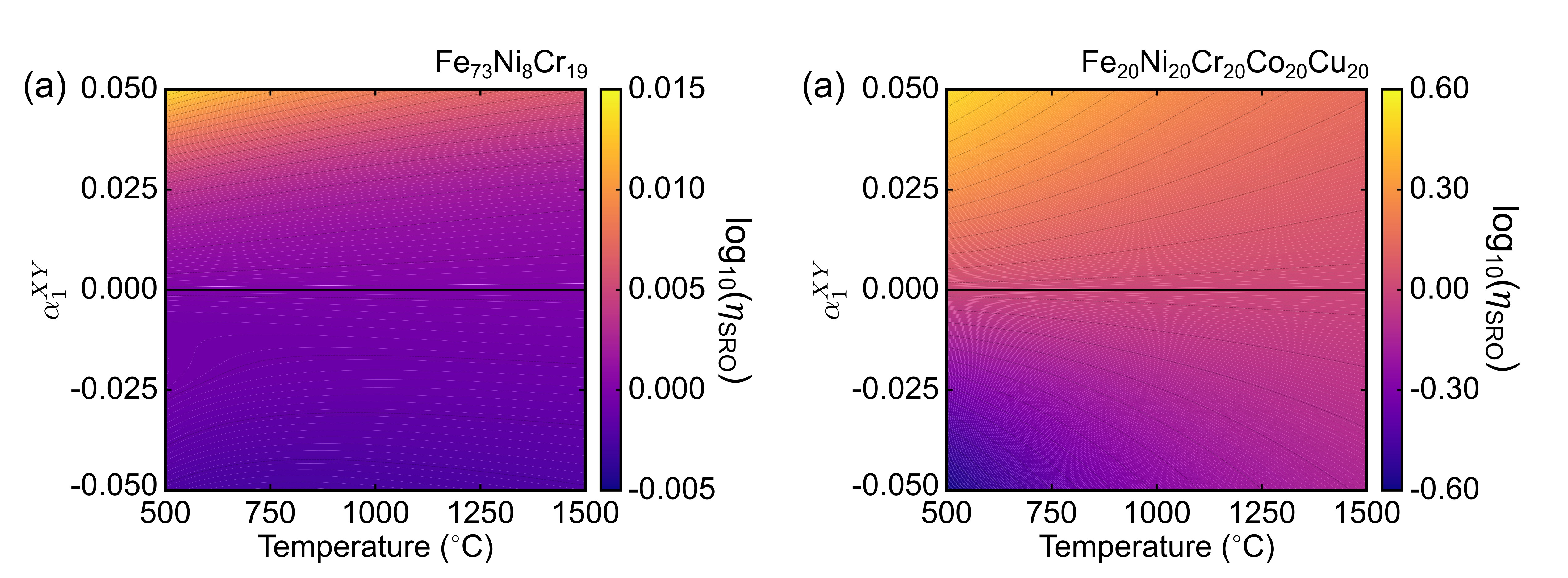}
    \caption{The predictions of property enhancement factors due to SRO. The enhancement factor is plotted for (a) the equilibrium vacancy concentration for Fe$_{73}$Ni$_{8}$Cr$_{19}$ and (b) the vacancy diffusivity for equimolar FeNiCrCoCu. Property enhancement factors are measured relative to the values for random arrangement (i.e., $\alpha_{\zeta}^{XY} = 0$). We note that predictions in (a) where temperature exceeds 1400~$^\circ$C are theoretical as they fall above the solidus temperature.}
    \label{fig:alphaContours}
\end{figure}

To explore the influence of SRO on vacancy-mediated properties, we calculate the enhancement factors for select vacancy properties and CSSs considered in this study using the statistics predicted by the mechanistic model. Here, the enhancement factors are calculated as each property evaluated with its effective barrier, normalized by its value under random arrangement. The lattice parameters and correlation factors are assumed to remain constant across all SRO configurations, such that all pre-exponential factors remain constant in Eqs.~\ref{eq:C_v} and~\ref{eq:D_v}. Under these considerations, the enhancement factors under SRO ($\eta_{SRO}$) are equivalently defined for vacancy concentration and vacancy diffusivity as:
\begin{equation}
\label{eq:enhance_SRO}
    \eta_{SRO} \equiv \left(\frac{\hat{C}_v\left(\alpha\right)}{\hat{C}_v\left(0\right)}\  \mathrm{or}\ \frac{\hat{D}_v\left(\alpha\right)}{\hat{D}_v\left(0\right)}\right) = \exp\left\{\frac{\hat{E}\left(0\right) - \hat{E}\left(\alpha\right)}{kT}\right\}
\end{equation}
where $\hat{E}\left(\alpha\right)$ and $\hat{E}\left(0\right)$ are the effective barriers for the SRO and random conditions, respectively, as obtained from the vacancy formation or migration energy distributions. Figure~\ref{fig:alphaContours}a provides the temperature-resolved vacancy concentration enhancement for Fe$_{73}$Ni$_{8}$Cr$_{19}$. As shown in the figure, this property exhibits a weak dependence on SRO across all conditions mapped, with enhancements capped at $\sim$3\%. The enhancement factors are distributed asymmetrically around the randomly arranged benchmark, with property increases observed for heteroatomic repulsion (i.e., $\alpha_1^{XY} > 0$) and property decreases for heteroatomic attraction (i.e., $\alpha_1^{XY} < 0$). Figure~\ref{fig:alphaContours}b provides the temperature-resolved vacancy diffusivity enhancement for equimolar FeNiCrCoCu. Here, the CSS exhibits a much stronger sensitivity to SRO, ranging from about a threefold increase at $\alpha_1^{XY} = 0.05$ to a reduction to roughly one quarter at $\alpha_1^{XY} = -0.05$. In both calculations, the enhancement factors are most significant at lower temperatures where thermal activation is suppressed, which is consistent with previous understandings from comparisons between mean and effective barriers under random arrangement.

\section{Conclusions}
This study presents a mechanistic model to predict the statistical parameters of vacancy energy distributions in CSSs with SRO. The model consists of a series of closed-form expressions derived from a statistical analysis of the salient excess energies that contribute to vacancy formation and migration processes. The model is free from adjustable parameters and is broadly applicable to systems with an arbitrary number of components, composition, and degrees of SRO, provided the atomic interactions can be treated by a pair-functional potential and the vacancy defect can be parameterized by a series of structure factors. Although several approximations are taken in the derivation and some limitations have been identified, the accuracy of predictions appears acceptable when compared to the resource-intensive methods used in common practice. In contrast to these methods, the model does not require computationally expensive simulation campaigns or large curated datasets to populate the defect energy space or train prediction models. Rather, the essential metallurgy is captured by the mechanistic model relations directly, which affords a dramatic speed-up in compute times and enables high-throughput screening activities. For all systems examined in this study, the statistical parameters for each configuration were obtained in $\sim$1~s using a modern desktop computer equipped with modest hardware. This model can therefore be leveraged to accelerate design efforts towards alloys with tailored diffusional flow characteristics and improved alloy performance.

Upon validation of the underlying relations of the mechanistic model, the vacancy energy distributions were studied for four different CSSs, two of which were also examined at different levels of SRO. Here, the model relations predicted the mean vacancy energies with excellent accuracy. The predictions for vacancy energy spreads were also in good agreement with direct simulations in most cases. However, the accuracy of VME standard deviation predictions was shown to degrade for systems where there were large separations in the sub-distributions of species-resolved vacancy properties. This discrepancy arises because the model does not resolve the identity of the migrating atom, and therefore omits the inter-species component of the spread. One significant finding is that maximizing configurational entropy is not a prerequisite to increasing vacancy energy fluctuations. This observation builds upon recent studies that have placed similar caveats on the role of configurational entropy in increasing the dispersion of defect energies.

To study the effects of fluctuations on vacancy-mediated properties, an effective barrier model was applied to predict the temperature dependence of the equilibrium vacancy concentration and vacancy diffusivity in the studied CSSs. These effective barrier calculations were compared with analogous mean barrier property calculations and validated against direct simulations and literature data, where available. Here, one important finding is that the mean barriers underpredicted the vacancy-mediated properties in all cases. This underprediction scales with the spread in the underlying vacancy energy distributions and is further exacerbated at intermediate and lower temperatures where thermal activation is suppressed. In the equimolar quinary alloy, where the vacancy energy spreads were largest, shortfalls reached $\sim$$10^3$ in the vacancy diffusivity and $\sim$$3 \times 10^2$ in the equilibrium vacancy concentration at the lower temperature extremes. The implications of fluctuations on the vacancy-mediated properties were further mapped by the mechanistic model across wide ranges of statistical parameters to explore sensitivities to standard deviations. The model was also used to examine the effects of SRO on vacancy-mediated properties. Here, a key finding is that SRO can have a meaningful impact, with changes of about threefold predicted under modest levels of SRO. Although this property change is relatively small compared to the overall fluctuation effect present across all configurations within the same CSS, it is nonetheless one that can arguably be controlled by materials processing and is therefore of interest to the community.

\appendix
\section{Derivation of the mean vacancy formation and migration energy expressions}
\label{AA}

This appendix derives the expressions for the mean VFE and VME in a form consistent with the structure-factor-based framework introduced in Section~\ref{vacancytheory}. In this section, these defect energies are expressed as the aggregation of excess energies from atoms within the cutoff of the vacancy defect. Here, we show how these  expressions may be obtained from their common definitions. The vacancy formation energy associated with the removal of an atom is defined as~\cite{24,27}:
\begin{equation}
E_{VFE} = E_{f} - E_{FCC} + \mu,
\label{eq:Append_VFE}
\end{equation}
where $E_{FCC}$ is the total energy of the perfect FCC crystal containing $N$ atoms, $E_{f}$ is the total energy of the monovacancy system with $(N-1)$ atoms, and $\mu$ is the chemical potential of the removed atom, accounting for the energy of the extracted atom. The total system energies may be obtained by summing over all atoms, such that Eq.~\ref{eq:Append_VFE} may be rewritten as:
\begin{equation}
E_{VFE} = \sum_{i=1}^{N-1}{e_{f}(i)} - \sum_{i=1}^{N}{e_{FCC}(i)} + \mu
\label{eq:Append_VFE2}
\end{equation}
where $e_{f}(i)$ and $e_{FCC}(i)$ are the cohesive energies of individual atoms in the monovacancy and FCC systems, respectively. For simplicity of notation, we often drop the $(i)$ index, but $e_{f}$ and $e_{FCC}$ retain this site-specific meaning throughout these derivations. The summations in Eq.~\ref{eq:Append_VFE2} may be further expanded by differentiating the $N_f$ sites that are within the cutoff of the vacancy from the $\left(N-1\right)-N_f$ sites that remain FCC coordinated. Furthermore, the $N^{\mathrm{th}}$ atom in the FCC system can be extracted from its summation, yielding the following expression:
\begin{subequations}
\begin{align}
E_{VFE} &= \sum_1^{N_f}{e_{f}} +\sum_{N_f+1}^{N-1}{e_{FCC}} - \sum_{1}^{N_f}{e_{FCC}} - \sum_{N_f+1}^{N-1}{e_{FCC}} - e_{FCC}+ \mu\\
&= \sum_{1}^{N_f}{\left(e_{f} - e_{FCC}\right)} - e_{FCC} + \mu \label{eq:Append_VFE3b}
\end{align}
\end{subequations}
Taking the expected value of Eq.~\ref{eq:Append_VFE3b}, and recognizing that $\langle e_{FCC}\rangle = \langle\mu\rangle$, we arrive at the expression:
\begin{subequations}
\begin{align}
\langle E_{VFE}\rangle & = \sum_{1}^{N_f}{\langle\left(e_{f} - e_{FCC}\right)\rangle} - \langle e_{FCC} \rangle + \langle{\mu}\rangle\\
\bar{E}_{VFE} &= \sum_{1}^{N_f}{\left(\bar{e}_{f} - \bar{e}_{FCC}\right)}
\label{eq:Append_VFE4b}
\end{align}
\end{subequations}
Here, $\bar{e}_{FCC}$ becomes a constant representing the mean cohesive energy of an FCC-coordinated atom in the CSS. Meanwhile, $\bar{e}_{f}$ remains site-specific and represents the different mean cohesive energies of the $N_f$ atoms that interact with the vacancy within the defect cutoff distance. Eq.~\ref{eq:Append_VFE4b} may be rewritten with structure factors by distributing the $N_f$ atoms into their corresponding neighbor shells around the monovacancy, resulting in the relation:
\begin{equation}
\bar{E}_{VFE} = \sum_{\zeta=1}^{\zeta_{cut}^f}Z_{\zeta}^{f}{\left(\bar{e}_{\zeta}^{f} - \bar{e}_{FCC}\right)}
\label{eq:Append_VFE5}
\end{equation}
which is the expression that appears in Section~\ref{vacancytheory} for the mean vacancy formation energy. 

By definition, the vacancy migration energy is taken as the difference between the system energy at the migration saddle point ($E_{s}$) and the monovacancy system energy. An excess energy-based relation for the mean vacancy migration energy may be obtained using a similar approach to that of the VFE, with separate accounting for atoms at defective sites around the migrating vacancy. That is, the system energies can be assembled from summations of all atomic sites, and the defective sites may be separated from this aggregate to form excess energy terms. These steps of the derivation proceed as follows:
\begin{subequations}
\begin{align}
E_{VME}
&= E_{s} - E_{f} \\
&= \left(E_{s} - E_{FCC}\right) - \left(E_{f} - E_{FCC}\right) \\
&= \left(\sum_{1}^{N-1} e_{s} - \sum_{1}^{N} e_{FCC}\right)
 - \left(\sum_{1}^{N-1} e_{f} - \sum_{1}^{N} e_{FCC}\right) \\
\begin{split}
&= \left(\sum_{1}^{N_s} e_{s} + \sum_{N_s+1}^{N-1} e_{FCC}
        - \sum_{1}^{N_s} e_{FCC} - \sum_{N_s+1}^{N-1} e_{FCC}
        - e_{FCC}\right) \\
&\quad - \left(\sum_{1}^{N_f} e_{f} + \sum_{N_f+1}^{N-1} e_{FCC}
        - \sum_{1}^{N_f} e_{FCC} - \sum_{N_f+1}^{N-1} e_{FCC}
        - e_{FCC}\right)
\end{split} \\
&= \sum_{1}^{N_s}\left(e_{s} - e_{FCC}\right)
 - \sum_{1}^{N_f}\left(e_{f} - e_{FCC}\right) \label{eq:Append_VME}
\end{align}
\end{subequations}
where $e_{s}$ are the cohesive energies of individual atoms in the saddle-point system and $N_{s}$ is the number of sites within the cutoff radius of the migrating vacancy. $e_{s}$ is analogous to $e_{f}$. We have again dropped the $(i)$ index and it retains the same site-specific definition when extracting its mean. Taking the expected value of Eq.~\ref{eq:Append_VME} and recognizing that $\sum^{N_f}\langle\left(e_{f} - e_{FCC}\right)\rangle = \bar{E}_{VFE}$, as in Eq.~\ref{eq:Append_VFE4b}, the relation reduces to:
\begin{equation}
\bar{E}_{VME} = \sum_{1}^{N_s}{\left(\bar{e}_{s} - \bar{e}_{FCC}\right)} - \bar{E}_{VFE}
\label{eq:Append_VME2}
\end{equation}
We may again express the VME relation using structure factors as:
\begin{equation}
\bar{E}_{VME} = \sum_{\zeta=1}^{\zeta_{cut}^{s}}{Z_{\zeta}^s\left(\bar{e}_{\zeta}^{s} - \bar{e}_{FCC}\right)} - \bar{E}_{VFE}
\label{eq:Append_VME3}
\end{equation}
which is the expression that appears in Section~\ref{vacancytheory} for the mean vacancy migration energy. Together with the derivation of $\bar{E}_{VFE}$, this Appendix demonstrates how vacancy formation and migration energies can be written as an aggregation of excess energies that are defined by the structure factors of the defective and reference crystal environment.

\section{Spread in vacancy formation and migration energies}
\label{AB}

This appendix provides a derivation of the expressions for the standard deviations of the VFE and VME that are introduced in Section~\ref{vacancytheory}. Our approach centers on the distribution of excess energies for sites within the cutoff radius of the vacancy defect. For the VFE, these individual site excess energies are written as $\Delta f(i) = e_{f}(i) - e_{FCC}(i)$, which is the difference of the cohesive energies for atoms in the monovacancy and FCC systems, respectively. Dropping the site-specific indices $(i)$, the standard deviations of these site excess energies ($\sigma_{\Delta f}$) may be obtained from the expected value formulation:
\begin{subequations}
\label{eq:AppendB_VFE_std_all}
\begin{align}
\sigma_{\Delta f}^{2} &= \langle \left[\left(e_{f} - e_{FCC}\right) - \left(\bar{e}_{f}-\bar{e}_{FCC}\right)\right]^{2}\rangle\\
&= \langle \left[\left(e_{f} - \bar{e}_{f}\right) - \left({e}_{FCC}-\bar{e}_{FCC}\right)\right]^{2}\rangle\\
&= \langle \left(e_{f} - \bar{e}_{f}\right)^2 -2\left(e_{f}-\bar{e}_{f}\right)\left(e_{FCC}-\bar{e}_{FCC}\right)+ \left({e}_{FCC}-\bar{e}_{FCC}\right)^{2}\rangle\\
&=\langle\left(e_{f} - \bar{e}_{f}\right)^2\rangle - 2\langle\left(e_{f}-\bar{e}_{f}\right)\left(e_{FCC}-\bar{e}_{FCC}\right)\rangle + \langle\left({e}_{FCC}-\bar{e}_{FCC}\right)^{2}\rangle\\
&=\sigma_{f}^2 - 2\mathrm{cov}\left(e_{f},e_{FCC}\right) + \sigma_{FCC}^2 \label{eq:AppendB_VFE_std}
\end{align}
\end{subequations}
where $\sigma_{f}$ and $\sigma_{FCC}$ are the standard deviations of the cohesive energies for the same site in the monovacancy and FCC crystal systems, respectively. Eq.~\ref{eq:AppendB_VFE_std} can be further simplified by assuming that the vacancy and FCC site energies are perfectly correlated such that $\mathrm{cov}\left(e_{f},e_{FCC}\right) = \sigma_{f}\sigma_{FCC}$: 
\begin{equation}
\label{eq:AppendB_VFE_std2}
\sigma_{\Delta f}^{2} = \sigma_{f}^2  + \sigma_{FCC}^2  - 2\sigma_{f}\sigma_{FCC}
\end{equation}
Each of the terms in Eq.~\ref{eq:AppendB_VFE_std2} can be obtained from the cohesive energy relations of Section~\ref{Theory}, with the explicit calculation differentiated only by their structure factors. With the expression for the spread of site excess energies fully developed, we now begin aggregating their contributions using the standard relations for addition of statistical distributions. For this calculation, we assume that site excess energies in the same neighbor shell around the monovacancy are perfectly correlated, but that there is no correlation between sites in different neighbor shells. It is helpful to collect the site excess energies into a single term ($\sigma_{\Delta f_{\zeta}}$), such that the aggregate contributions of site excess energies to the spread become $\sum_{\zeta}Z_{\zeta}^f \sigma_{\Delta f_{\zeta}}^2$. Furthermore, the site excess energies may be separately enumerated and compared as $\sigma_{\Delta f}(i)$ and $\sigma_{\Delta f}(j)$, where $i$ and $j$ refer to specific sites from the $N_f$ within the cutoff radius. Under these correlation assumptions $\mathrm{cov}\left(\Delta f(i),\Delta f(j)\right) = \sigma_{\Delta f}(i)\sigma_{\Delta f}(j) = \sigma_{\Delta f_{\zeta}}^2$ when $i$ and $j$ refer to sites within the same neighbor shell, but is zero for all other pairings. Consequently, the standard deviation of the vacancy formation energy ($\sigma_{VFE}$) is obtained from:
\begin{subequations}
\begin{align}
\sigma_{VFE}^{2} &= \sum_{\zeta=1}^{\zeta_{cut}^{f}}Z_{\zeta}^f\sigma_{\Delta f_{\zeta}}^2 + 2\sum_{i<j}^{N_f}\mathrm{cov}\left(\Delta f(i),\Delta f(j)\right)\\
&= \sum_{\zeta=1}^{\zeta_{cut}^{f}}Z_{\zeta}^f\sigma_{\Delta f_{\zeta}}^2 + \sum_{\zeta = 1}^{\zeta_{cut}^f}Z_{\zeta}^f\left(Z_{\zeta}^f-1\right)\sigma_{\Delta f_{\zeta}}^2\\
&= \sum_{\zeta=1}^{\zeta_{cut}^{f}}\left(Z_{\zeta}^f\sigma_{\Delta f_{\zeta}}\right)^2 \label{eq:AppendB_VFE_std3}
\end{align}
\end{subequations}
which is Eq.~\ref{eq:VFE_std} from Section~\ref{vacancytheory}. We note that this treatment of the standard deviation neglects the variance introduced by differences between the chemical potential of the removed species ($\mu^X$) and its cohesive energy in solution ($e^X_{FCC}$), $\mathrm{Var}\left[\mu^X - e^X_{FCC}\right]$, which is included in some approaches~\cite{27}. Here, $\mathrm{Var}[\,\cdot\,]$ denotes the variance operator. This simplification removes the need to calculate the chemical potential in each SRO configuration, at the cost of only a small change to the predicted equilibrium vacancy concentrations (see Figure~\ref{fig:Prop}).

The standard deviation of the vacancy migration energies may be calculated following a similar site excess energy approach. For the VME, the individual site excess energies $e_{s}(i) - e_{FCC}(i)$ are of interest, which represent the difference of cohesive energies for atoms in the saddle-point and FCC systems, respectively. Under the same operations and assumptions of Eq.~\ref{eq:AppendB_VFE_std_all}, the standard deviations of the site excess energies ($\sigma_{\Delta s}$) are obtained from:
\begin{equation}
\label{eq:AppendB_VME_std}
\sigma_{\Delta s}^{2} = \sigma_{s}^2  + \sigma_{FCC}^2  - 2\sigma_{s}\sigma_{FCC}
\end{equation}
where $\sigma_{s}$ is the standard deviation of the cohesive energies for the same site in the saddle-point configuration. As with the vacancy formation process, these values may be collected within shared neighbor shells and written as a standard deviation resolved by structure factor ($\sigma_{\Delta s_{\zeta}}$), albeit with a greater number of sites within the cutoff radius ($N_s > N_f$). Following the same intra-shell and inter-shell correlation assumptions from the VFE analysis, the contributions of the spreads in site excess energies may be aggregated as $\sum_{\zeta}\left(Z_{\zeta}^{s}\sigma_{\Delta s_{\zeta}}\right)^2$. In its current form, the expression for site excess energies is referenced against the FCC crystal, but the VME is measured relative to the monovacancy configuration. This mismatch may be reconciled through examination of the excess energy expressions of~\ref{AA}. Here, we have shown how the vacancy migration energy can be obtained from summation of site excess cohesive energies and subtraction of the VFE baseline: $E_{VME} = \sum_{\zeta}Z_{\zeta}^{s}\left(e_{s} - e_{FCC}\right) - E_{VFE}$. We note that the spread in the first term of this expression is $\sum_{\zeta}\left(Z_{\zeta}^{s}\sigma_{\Delta s_{\zeta}}\right)^2$ and that the spread in the second term is $\sigma_{VFE}^2$. Assuming that the formation and saddle excess energies are uncorrelated, the standard deviation of the vacancy migration energy ($\sigma_{VME}$) is obtained by taking the variance of this VME expression:
\begin{subequations}
\begin{align}
\mathrm{Var}\left[E_{VME}\right]  &= \mathrm{Var}\left[\sum_{\zeta=1}^{\zeta_{cut}^{s}}Z_{\zeta}^{s}\left(e_{s} - e_{FCC}\right) - E_{VFE}\right]\\
\sigma_{VME}^2 &= \mathrm{Var}\left[\sum_{\zeta=1}^{\zeta_{cut}^{s}}Z_{\zeta}^{s}\left(e_{s} - e_{FCC}\right)\right] + \mathrm{Var}\left[E_{VFE}\right]\\
&= \sum_{\zeta=1}^{\zeta_{cut}^{s}}\left(Z_{\zeta}^s\sigma_{\Delta s_{\zeta}}\right)^2 + \sigma_{VFE}^2 \label{eq:AppendB_VME_std2}
\end{align}
\end{subequations}
This expression for the standard deviation of the vacancy migration energy appears as Eq.~\ref{eq:VME_std} from Section~\ref{vacancytheory}. 

\section*{Acknowledgments}
This material is based upon work supported by the National Science Foundation under Grant No. DMR-2144451. The authors gratefully acknowledge the Advanced Cyberinfrastructure for Education and Research (ACER) group at the University of Illinois Chicago for providing the computational resources and services required to present the research results in this paper.

\section*{Declaration of generative AI and AI-assisted technologies in the manuscript preparation process}
During the preparation of this work the author(s) used Anthropic's Claude Opus 5.5 model to double-check the authors' algebra in model derivations, verify figure data against codes, check equation typesetting, and cross-check in-text referencing. After using this tool/service, the author(s) reviewed and edited the content as needed and take(s) full responsibility for the content of the published article.

\bibliographystyle{elsarticle-num} 
\bibliography{references.bib}

@article{Zhao2017,
author = {Zhao, Shijun and Osetsky, Yuri and Zhang, Yanwen},
doi = {10.1016/j.actamat.2017.01.056},
issn = {13596454},
journal = {Acta Materialia},
month = {apr},
pages = {391--399},
title = {{Preferential diffusion in concentrated solid solution alloys: NiFe, NiCo and NiCoCr}},
volume = {128},
year = {2017}
}

@article{Osetsky2016,
author = {Osetsky, Yuri N. and B{\'{e}}land, Laurent K. and Stoller, Roger E.},
doi = {10.1016/j.actamat.2016.06.018},
issn = {13596454},
journal = {Acta Materialia},
month = {aug},
pages = {364--371},
title = {{Specific features of defect and mass transport in concentrated fcc alloys}},
volume = {115},
year = {2016}
}

@article{23,
  title={Defect energetics of concentrated solid-solution alloys from ab initio calculations: {Ni$_{0.5}$Co$_{0.5}$}, {Ni$_{0.5}$Fe$_{0.5}$}, {Ni$_{0.8}$Fe$_{0.2}$} and {Ni$_{0.8}$Cr$_{0.2}$}},
  author={Zhao, Shijun and Stocks, G Malcolm and Zhang, Yanwen},
  journal={Physical Chemistry Chemical Physics},
  volume={18},
  number={34},
  pages={24043--24056},
  year={2016},
  publisher={Royal Society of Chemistry}
}

@article{24,
  title={Vacancy formation energies and migration barriers in multi-principal element alloys},
  author={Roy, Ankit and Singh, Prashant and Balasubramanian, Ganesh and Johnson, Duane D},
  journal={Acta Materialia},
  volume={226},
  pages={117611},
  year={2022},
  publisher={Elsevier}
}

@article{27,
  title={A statistical approach for atomistic calculations of vacancy formation energy and chemical potentials in concentrated solid-solution alloys},
  author={Zhang, Yongfeng and Manzoor, Anus and Jiang, Chao and Aidhy, Dilpuneet and Schwen, Daniel},
  journal={Computational Materials Science},
  volume={190},
  pages={110308},
  year={2021},
  publisher={Elsevier}
}

@article{32,
author = {An, Xudong and Lu, Eryang and Makkonen, Ilja and Wei, Guanying and Byggm{\"{a}}star, Jesper and Zhu, Jiulong and Mizohata, Kenichiro and Chen, Zhehao and Djurabekova, Flyura and Hu, Wangyu and Deng, Huiqiu and Yang, Tengfei and Tuomisto, Filip},
doi = {10.1016/j.jallcom.2024.177704},
issn = {09258388},
journal = {Journal of Alloys and Compounds},
month = {jan},
pages = {177704},
publisher = {Elsevier},
title={Enhanced migration of mono-vacancies in {Al$_x$FeCoCrNi} high entropy alloys},
volume = {1010},
year = {2025}
}

@article{34,
   title={Thermodynamic mixing energy and heterogeneous diffusion uncover the mechanisms of radiation damage reduction in single-phase {Ni-Fe} alloys},
  author={Jin, Miaomiao and Cao, Penghui and Short, Michael P},
  journal={Acta Materialia},
  volume={147},
  pages={16--23},
  year={2018},
  publisher={Elsevier}
}

@article{35,
  title={Modeling environment-dependent atomic-level properties in complex-concentrated alloys},
  author={Farnell, Mackinzie S and McClure, Zachary D and Tripathi, Shivam and Strachan, Alejandro},
  journal={The Journal of Chemical Physics},
  volume={156},
  number={11},
  pages={114102},
  year={2022},
  publisher={AIP Publishing}
}

@article{plimpton1995fast,
  title={Fast parallel algorithms for short-range molecular dynamics},
  author={Plimpton, Steve},
  journal={Journal of Computational Physics},
  volume={117},
  number={1},
  pages={1--19},
  year={1995},
  publisher={Elsevier}
}

@article{varvenne2016average,
  title={Average-atom interatomic potential for random alloys},
  author={Varvenne, C{\'e}line and Luque, Aitor and N{\"o}hring, Wolfram G and Curtin, William A},
  journal={Physical Review B},
  volume={93},
  number={10},
  pages={104201},
  year={2016},
  publisher={APS}
}

@article{38,
  title={The embedded-atom method: a review of theory and applications},
  author={Daw, Murray S and Foiles, Stephen M and Baskes, Michael I},
  journal={Materials Science Reports},
  volume={9},
  number={7-8},
  pages={251--310},
  year={1993},
  publisher={Elsevier}
}

@article{39,
  title={Embedded-atom-method functions for the {fcc} metals {Cu}, {Ag}, {Au}, {Ni}, {Pd}, {Pt}, and their alloys},
  author={Foiles, S. M. and Baskes, M. I. and Daw, Murray S},
  journal={Physical Review B},
  volume={33},
  number={12},
  pages={7983},
  year={1986},
  publisher={APS}
}

@article{40,
  title={Embedded-atom method: Derivation and application to impurities, surfaces, and other defects in metals},
  author={Daw, Murray S and Baskes, Michael I},
  journal={Physical Review B},
  volume={29},
  number={12},
  pages={6443},
  year={1984},
  publisher={APS}
}

@article{jagatramka2022analytical,
  title={An analytical method to quantify the statistics of energy landscapes in random solid solutions},
  author={Jagatramka, Ritesh and Wang, Chu and Daly, Matthew},
  journal={Computational Materials Science},
  volume={214},
  pages={111763},
  year={2022},
  publisher={Elsevier}
}

@article{stukowski2009visualization,
  title={Visualization and analysis of atomistic simulation data with {OVITO}--the {Open Visualization Tool}},
  author={Stukowski, Alexander},
  journal={Modelling and Simulation in Materials Science and Engineering},
  volume={18},
  number={1},
  pages={015012},
  year={2010},
  publisher={IOP Publishing}
}

@article{del2011formation,
  title={Formation energy of vacancies in {FeCr} alloys: Dependence on {Cr} concentration},
  author={Del Rio, Emma and Sampedro, Jes{\'u}s M and Dogo, Harun and Caturla, Mar{\'\i}a J and Caro, Magdalena and Caro, Alfredo and Perlado, J Manuel},
  journal={Journal of Nuclear Materials},
  volume={408},
  number={1},
  pages={18--24},
  year={2011},
  publisher={Elsevier}
}

@article{wang2018energetics,
  title={Energetics analysis of interstitial loops in single-phase concentrated solid-solution alloys},
  author={Wang, Xin-Xin and Niu, Liang-Liang and Wang, Shaoqing},
  journal={Journal of Nuclear Materials},
  volume={501},
  pages={94--103},
  year={2018},
  publisher={Elsevier}
}

@article{deluigi2021simulations,
  title={Simulations of primary damage in a High Entropy Alloy: Probing enhanced radiation resistance},
  author={Deluigi, Orlando Raul and Pasianot, Roberto Cesar and Valencia, F. J. and Caro, A. and Farkas, D. and Bringa, Eduardo Marcial},
  journal={Acta Materialia},
  volume={213},
  pages={116951},
  year={2021},
  publisher={Elsevier}
}

@article{RD3,
  title={Interatomic potential to study plasticity in stainless steels: the {FeNiCr} model alloy},
  author={Bonny, G and Terentyev, D and Pasianot, Roberto Cesar and Ponc{\'e}, Samuel and Bakaev, Alexander},
  journal={Modelling and Simulation in Materials Science and Engineering},
  volume={19},
  number={8},
  pages={085008},
  year={2011},
  publisher={IOP Publishing}
}

@article{zhang2022ab,
  title={Ab initio prediction of vacancy energetics in {HCP} {Al-Hf-Sc-Ti-Zr} high entropy alloys and the subsystems},
  author={Zhang, Xi and Divinski, Sergiy V and Grabowski, Blazej},
  journal={Acta Materialia},
  volume={227},
  pages={117677},
  year={2022},
  publisher={Elsevier}
}

@article{zhao2017suppression,
  title={Suppression of vacancy cluster growth in concentrated solid solution alloys},
  author={Zhao, Shijun and Velisa, Gihan and Xue, Haizhou and Bei, Hongbin and Weber, William J and Zhang, Yanwen},
  journal={Acta Materialia},
  volume={125},
  pages={231--237},
  year={2017},
  publisher={Elsevier}
}

@article{jagatramka2023evolution,
  title={The evolution of deformation twinning microstructures in random face-centered cubic solid solutions},
  author={Jagatramka, Ritesh and Ahmed, Junaid and Daly, Matthew},
  journal={Journal of Applied Physics},
  volume={133},
  number={5},
  pages={055107},
  year={2023},
  publisher={AIP Publishing}
}

@article{rasooli2024deformation,
  title={Deformation mechanisms in high entropy alloys: a minireview of short-range order effects},
  author={Rasooli, Novin and Chen, Wei and Daly, Matthew},
  journal={Nanoscale},
  volume={16},
  number={4},
  pages={1650--1663},
  year={2024},
  publisher={Royal Society of Chemistry}
}

@article{gao2024defect,
   title={Defect energetics in an high-entropy alloy {fcc} {CoCrFeMnNi}},
  author={Gao, Chan and Wang, Shuyu and Liu, Xiao and Singh, Chandra Veer},
  journal={Materials Advances},
  volume={5},
  number={10},
  pages={4231--4241},
  year={2024},
  publisher={Royal Society of Chemistry}
}

@article{zhao2020defect,
  title={Defect properties in a {VTaCrW} equiatomic high entropy alloy ({HEA}) with the body centered cubic ({bcc}) structure},
  author={Zhao, Shijun},
  journal={Journal of Materials Science \& Technology},
  volume={44},
  pages={133--139},
  year={2020},
  publisher={Elsevier}
}

@article{li2019first,
  title={First principle study of magnetism and vacancy energetics in a near equimolar {NiFeMnCr} high entropy alloy},
  author={Li, Congyi and Yin, Junqi and Odbadrakh, Khorgolkhuu and Sales, Brian C and Zinkle, Steven J and Stocks, G Malcolm and Wirth, Brian D},
  journal={Journal of Applied Physics},
  volume={125},
  number={15},
  year={2019},
  pages={155103},
  publisher={AIP Publishing}
}

@article{xu2023mechanism,
  title={Mechanism of sluggish diffusion under rough energy landscape},
  author={Xu, Biao and Zhang, Jun and Xiong, Yaoxu and Ma, Shihua and Osetsky, Yuri and Zhao, Shijun},
  journal={Cell Reports Physical Science},
  volume={4},
  number={4},
  pages={101337},
  year={2023},
  publisher={Elsevier}
}

@article{xu2023influence,
  title={Influence of short-range order on diffusion in multiprincipal element alloys from long-time atomistic simulations},
  author={Xu, Biao and Ma, Shihua and Huang, Shasha and Zhang, Jun and Xiong, Yaoxu and Fu, Haijun and Xiang, Xuepeng and Zhao, Shijun},
  journal={Physical Review Materials},
  volume={7},
  number={3},
  pages={033605},
  year={2023},
  publisher={APS}
}

@article{xing2024neural,
  title={Neural network kinetics for exploring diffusion multiplicity and chemical ordering in compositionally complex materials},
  author={Xing, Bin and Rupert, Timothy J and Pan, Xiaoqing and Cao, Penghui},
  journal={Nature Communications},
  volume={15},
  number={1},
  pages={3879},
  year={2024},
  publisher={Nature Publishing Group UK London}
}

@article{xing2022short,
  title={Short-range order localizing diffusion in multi-principal element alloys},
  author={Xing, Bin and Wang, Xinyi and Bowman, William J and Cao, Penghui},
  journal={Scripta Materialia},
  volume={210},
  pages={114450},
  year={2022},
  publisher={Elsevier}
}

@article{kang2023density,
  title={Density functional study of atomic arrangements in {CrMnFeCoNi} high-entropy alloy and their impact on vacancy formation energy and segregation},
  author={Kang, ShinYoung and Tamm, Artur},
  journal={Computational Materials Science},
  volume={230},
  pages={112456},
  year={2023},
  publisher={Elsevier}
}

@article{cao2021does,
  title={How does short-range order impact defect kinetics in irradiated multiprincipal element alloys?},
  author={Cao, Penghui},
  journal={Accounts of Materials Research},
  volume={2},
  number={2},
  pages={71--74},
  year={2021},
  publisher={ACS Publications}
}

@article{li2024vacancy,
  title={Vacancy formation free energy in concentrated alloys: Equilibrium vs. random sampling},
  author={Li, Kangming and Schuler, Thomas and Fu, Chu-Chun and Nastar, Maylise},
  journal={Acta Materialia},
  volume={281},
  pages={120355},
  year={2024},
  publisher={Elsevier}
}

@article{cowley1965short,
  title={Short-range order and long-range order parameters},
  author={Cowley, J. M.},
  journal={Physical Review},
  volume={138},
  number={5A},
  pages={A1384},
  year={1965},
  publisher={APS}
}

@article{xing2024vacancy,
  title={Vacancy diffusion barrier spectrum and diffusion correlation in multicomponent alloys},
  author={Xing, Bin and Zou, Wanjuan and Rupert, Timothy J and Cao, Penghui},
  journal={Acta Materialia},
  volume={266},
  pages={119653},
  year={2024},
  publisher={Elsevier}
}

@article{ibrahim2026machine,
  title={Machine learning prediction of vacancy formation energies in {CoNiCrFe} high-entropy alloy: The role of atomic descriptors and local chemical order},
  author={Ibrahim, Shehu Adam and Yang, Jinxue and Shi, Tan and Zhang, Chen and Chen, Da and Li, Jing and Li, Yang and Mbazor, Jeremiah Chinonso and Zhang, Yizhuo and Su, Zhengxiong and others},
  journal={The Journal of Physical Chemistry C},
  volume={130},
  number={12},
  pages={4538--4553},
  year={2026},
  publisher={ACS Publications}
}

@article{fan2022predicting,
  title={Predicting path-dependent diffusion barrier spectra in vast compositional space of multi-principal element alloys via convolutional neural networks},
  author={Fan, Zhao and Xing, Bin and Cao, Penghui},
  journal={Acta Materialia},
  volume={237},
  pages={118159},
  year={2022},
  publisher={Elsevier}
}

@article{castin2014predicting,
  title={Predicting vacancy migration energies in lattice-free environments using artificial neural networks},
  author={Castin, Nicolas and Fern{\'a}ndez, J. R. and Pasianot, Roberto Cesar},
  journal={Computational Materials Science},
  volume={84},
  pages={217--225},
  year={2014},
  publisher={Elsevier}
}

@article{choudhary2023can,
  title={Can a deep-learning model make fast predictions of vacancy formation in diverse materials?},
  author={Choudhary, Kamal and Sumpter, Bobby G},
  journal={AIP Advances},
  volume={13},
  number={9},
  pages={095109},
  year={2023},
  publisher={AIP Publishing}
}

@article{manzoor2024interplay,
  title={Interplay between thermal vacancy and short-range order in complex concentrated alloys},
  author={Manzoor, Anus and Zhang, Yongfeng},
  journal={Journal of Alloys and Compounds},
  volume={982},
  pages={173788},
  year={2024},
  publisher={Elsevier}
}

@article{miracle2017critical,
  title={A critical review of high entropy alloys and related concepts},
  author={Miracle, Daniel B and Senkov, Oleg N},
  journal={Acta Materialia},
  volume={122},
  pages={448--511},
  year={2017},
  publisher={Elsevier}
}

@article{george2020high,
  title={High entropy alloys: A focused review of mechanical properties and deformation mechanisms},
  author={George, Easo P and Curtin, William A and Tasan, Cemal Cem},
  journal={Acta Materialia},
  volume={188},
  pages={435--474},
  year={2020},
  publisher={Elsevier}
}

@article{george2019high,
  title={High-entropy alloys},
  author={George, Easo P and Raabe, Dierk and Ritchie, Robert O},
  journal={Nature Reviews Materials},
  volume={4},
  number={8},
  pages={515--534},
  year={2019},
  publisher={Nature Publishing Group UK London}
}

@article{zhang2014microstructures,
  title={Microstructures and properties of high-entropy alloys},
  author={Zhang, Yong and Zuo, Ting Ting and Tang, Zhi and Gao, Michael C and Dahmen, Karin A and Liaw, Peter K and Lu, Zhao Ping},
  journal={Progress in Materials Science},
  volume={61},
  pages={1--93},
  year={2014},
  publisher={Elsevier}
}

@article{Beland2016,
author = {B{\'{e}}land, Laurent Karim and Lu, Chenyang and Osetskiy, Yuri N. and Samolyuk, German D. and Caro, Alfredo and Wang, Lumin and Stoller, Roger E.},
doi = {10.1063/1.4942533},
journal = {Journal of Applied Physics},
month = {feb},
number = {8},
pages = {085901},
title = {{Features of primary damage by high energy displacement cascades in concentrated Ni-based alloys}},
volume = {119},
year = {2016}
}

@article{Osetsky2018,
author = {Osetsky, Yuri N. and B{\'{e}}land, Laurent K. and Barashev, Alexander V. and Zhang, Yanwen},
doi = {10.1016/j.cossms.2018.05.003},
issn = {13590286},
journal = {Current Opinion in Solid State and Materials Science},
month = {jun},
number = {3},
pages = {65--74},
title = {{On the existence and origin of sluggish diffusion in chemically disordered concentrated alloys}},
volume = {22},
year = {2018}
}

@article{LaRosa2022,
author = {LaRosa, Carlyn R. and Ghazisaeidi, Maryam},
doi = {10.1016/j.actamat.2022.118165},
journal = {Acta Materialia},
month = {oct},
pages = {118165},
publisher = {Elsevier Ltd},
title = {{A ``local'' stacking fault energy model for concentrated alloys}},
volume = {238},
year = {2022}
}

@article{Sheriff2024,
author = {Sheriff, Killian and Cao, Yifan and Smidt, Tess and Freitas, Rodrigo},
doi = {10.1073/pnas.2322962121},
journal = {Proceedings of the National Academy of Sciences},
month = {jun},
number = {25},
pages = {e2322962121},
pmid = {38870054},
title = {{Quantifying chemical short-range order in metallic alloys}},
volume = {121},
year = {2024}
}

@book{Mehrer2007,
address = {Berlin, Heidelberg},
author = {Mehrer, Helmut},
doi = {10.1007/978-3-540-71488-0},
isbn = {978-3-540-71486-6},
publisher = {Springer Berlin Heidelberg},
series = {Springer Series in Solid-State Sciences},
title = {{Diffusion in Solids}},
volume = {155},
year = {2007}
}

@article{Wang2022,
author = {Wang, Xinyi and Maresca, Francesco and Cao, Penghui},
doi = {10.1016/j.actamat.2022.118022},
issn = {13596454},
journal = {Acta Materialia},
publisher = {Elsevier Ltd},
title = {{The hierarchical energy landscape of screw dislocation motion in refractory high-entropy alloys}},
volume = {234},
pages = {118022},
year = {2022}
}

@article{manzoor2021machine,
  title={Machine learning based methodology to predict point defect energies in multi-principal element alloys},
  author={Manzoor, Anus and Arora, Gaurav and Jerome, Bryant and Linton, Nathan and Norman, Bailey and Aidhy, Dilpuneet S},
  journal={Frontiers in Materials},
  volume={8},
  pages={673574},
  year={2021},
  publisher={Frontiers Media SA}
}

@article{guo2024properties,
  title={Properties of radiation-induced point defects in austenitic steels: a molecular dynamics study},
  author={Guo, Wei and Liang, Yanxiang and Wan, Qiang},
  journal={Modelling and Simulation in Materials Science and Engineering},
  volume={32},
  number={5},
  pages={055022},
  year={2024},
  publisher={IOP Publishing}
}

@article{He2024,
author = {He, Mengwei and Davids, William J. and Breen, Andrew J. and Ringer, Simon P.},
doi = {10.1038/s41563-024-01912-1},
issn = {14764660},
journal = {Nature Materials},
number = {9},
pages = {1200--1207},
pmid = {38956352},
publisher = {Springer US},
title = {{Quantifying short-range order using atom probe tomography}},
volume = {23},
year = {2024}
}

@misc{AdvancedMaterialsandMicrostructuresLabattheUniversityofIllinoisChicago,
author = {{Advanced Materials and Microstructures Lab at the University of Illinois Chicago}},
title = {{Solid Solution PE Statistics}},
url = {https://github.com/dalymatt/Solid-Solution-PE-Statistics},
year={2026}
}

@article{du2022chemical,
  title={Chemical domain structure and its formation kinetics in {CrCoNi} medium-entropy alloy},
  author={Du, Jun-Ping and Yu, Peijun and Shinzato, Shuhei and Meng, Fan-Shun and Sato, Yuji and Li, Yangen and Fan, Yiwen and Ogata, Shigenobu},
  journal={Acta Materialia},
  volume={240},
  pages={118314},
  year={2022},
  publisher={Elsevier}
}

@article{burton1972vacancy,
  title={Vacancy-formation entropy in cubic metals},
  author={Burton, J. J.},
  journal={Physical Review B},
  volume={5},
  number={8},
  pages={2948},
  year={1972},
  publisher={APS}
}

@article{Klueh2007,
author = {Klueh, R. L. and Nelson, A. T.},
doi = {10.1016/j.jnucmat.2007.05.005},
issn = {00223115},
journal = {Journal of Nuclear Materials},
month = {sep},
number = {1-3},
pages = {37--52},
title = {{Ferritic/martensitic steels for next-generation reactors}},
volume = {371},
year = {2007}
}

@article{cowley1950x,
  title={X-ray measurement of order in single crystals of {Cu$_3$Au}},
  author={Cowley, J. M.},
  journal={Journal of Applied Physics},
  volume={21},
  number={1},
  pages={24--30},
  year={1950},
  publisher={American Institute of Physics}
}

@article{Thompson2022,
author = {Thompson, Aidan P. and Aktulga, H. Metin and Berger, Richard and Bolintineanu, Dan S. and Brown, W. Michael and Crozier, Paul S. and {in 't Veld}, Pieter J. and Kohlmeyer, Axel and Moore, Stan G. and Nguyen, Trung Dac and Shan, Ray and Stevens, Mark J. and Tranchida, Julien and Trott, Christian and Plimpton, Steven J.},
doi = {10.1016/j.cpc.2021.108171},
issn = {00104655},
journal = {Computer Physics Communications},
month = {feb},
pages = {108171},
title = {{LAMMPS -- a flexible simulation tool for particle-based materials modeling at the atomic, meso, and continuum scales}},
volume = {271},
year = {2022}
}

@article{Henkelman2000,
author = {Henkelman, Graeme and Uberuaga, Blas P. and J{\'{o}}nsson, Hannes},
doi = {10.1063/1.1329672},
issn = {0021-9606},
journal = {The Journal of Chemical Physics},
month = {dec},
number = {22},
pages = {9901--9904},
title = {{A climbing image nudged elastic band method for finding saddle points and minimum energy paths}},
volume = {113},
year = {2000}
}

@article{Lee2026,
author = {Lee, Damien K. J. and M{\"{u}}ller, Yann L. and Natarajan, Anirudh Raju},
doi = {10.1016/j.actamat.2025.121752},
issn = {13596454},
journal = {Acta Materialia},
month = {jan},
pages = {121752},
title = {{Modeling the equilibrium vacancy concentration in multi-principal element alloys from first-principles}},
volume = {304},
year = {2026}
}

@article{Ruban2016,
author = {Ruban, A. V.},
doi = {10.1103/PhysRevB.93.134115},
issn = {2469-9950},
journal = {Physical Review B},
month = {apr},
number = {13},
pages = {134115},
title = {{Thermal vacancies in random alloys in the single-site mean-field approximation}},
volume = {93},
year = {2016}
}
\end{document}